\documentclass[12pt]{article}
\usepackage[english]{babel}
\usepackage{multirow}
\usepackage{natbib}
\usepackage{amsmath,amsfonts,amssymb,amsthm}
\usepackage{longtable}
\usepackage{dsfont}
\usepackage[official]{eurosym}
\usepackage{graphicx,graphics}
\usepackage{booktabs}
\usepackage[flushleft]{threeparttable}
\usepackage{url}
\usepackage[breaklinks=true]{hyperref}
\usepackage{pdflscape,pdfpages}
\usepackage{xcolor,colortbl}
\usepackage{rotating}
\usepackage{booktabs}
\usepackage{dcolumn}
\usepackage{float}

\usepackage[justification=justified]{caption}
\usepackage{subcaption}
\usepackage{subfiles}
\usepackage[toc, page]{appendix}
\usepackage{comment}
\usepackage{bbm}
\usepackage{dirtytalk}

\usepackage[colorinlistoftodos]{todonotes}

\usepackage{color,soul}

\usepackage{pict2e}
\usepackage{tikz}
\usetikzlibrary{patterns}

\usepackage[colorinlistoftodos]{todonotes}

\hypersetup{colorlinks,linkcolor={red!50!black},citecolor={blue!50!black},urlcolor={blue!80!black}
}
 
\usepackage[sc]{mathpazo}
\usepackage[margin=1in]{geometry}

\usepackage{setspace} 
\usepackage{tabularx}
\newcolumntype{z}{>{\centering\arraybackslash}X}

\title{ \vspace*{-2.5cm} \hspace*{-0.5cm}The Heterogeneous Earnings Impact of Job Loss Across Workers, Establishments, and Markets\footnote{
We thank Stefan Eriksson, Peter Fredriksson, Vitor Hadad, Martin Huber, Michael Lechner, Nicolaj Mühlbach, Stefan Pitschner, Kjell Salvanes, David Strömberg, Erik Sverdrup and seminar participants at Aalto, Cattolica Milan, CREAM, EALE, IfFS, Ratio, SKILS, SOLE, SSE, St Andrews, Stanford HAI, Swedish Economics Meeting, Umeå, Uppsala and Örebro. Skans and Yakymovych were supported by Vetenskapsrådet (2018-04581). Skans was also supported by Riksbankens Jubileumsfond Grant M23-0019. Athey and Simon acknowledge support from the Golub Capital Social Impact Lab at Stanford GSB.
}}

\author{Susan Athey\thanks{Stanford University, GSB} \and Lisa K. Simon\thanks{Revelio Labs, New York} \and Oskar N. Skans\thanks{Dep. of Economics, Uppsala University oskar.nordstrom\_skans@nek.uu.se} \and Johan Vikström\thanks{IFAU; Dep. of Economics, Uppsala University} \and Yaroslav Yakymovych\thanks{Institute for Housing and Urban Research, Uppsala University }
}

\date{ \vspace*{0.5cm} \today} 

\usepackage{natbib}
\usepackage{graphicx}
\usepackage{xcolor}
\begin{document}

\thispagestyle{empty}
\maketitle
 \vspace{-2.75em}
\begin{abstract} 
\noindent
Using rich Swedish administrative data, we apply causal machine learning methods to study how earnings losses after job displacement vary with observable characteristics that may be relevant for targeting policy interventions for workers. Heterogeneity in effects is as large within as across worker groups defined by age and schooling, and as large within as across establishments. A substantial portion of cross-establishment heterogeneity can be explained by industry and local labor market characteristics, suggesting a role for place- and industry-based targeting. The largest losses are concentrated among already vulnerable workers, indicating that well-designed targeting policies can improve both efficiency and equity.
\textbf{JEL-codes}: J65, J21, J31, C45

\end{abstract}

\maketitle
\newpage

\section{Introduction}

\onehalfspacing
Structural change and job reallocation drive economic growth \citep{BartelsmanDoms} but impose large and persistent earnings losses on displaced workers \citep{JLS1993}.\footnote{See also, e.g., \cite{sullivan09} on mortality, \cite{Eliason2014} on increased drinking, \cite{Black} on increased smoking, and \cite{Eliason2012} on increased divorces.} Governments therefore devote substantial resources to mitigating the individual consequences of job destruction \citep{OECD_SaG_2019}. These efforts include both \textit{ex post} redistribution and retraining programs and \textit{ex ante} policies such as short-time work schemes, employment protection, and preventive training initiatives aimed at workers at risk. Because interventions are costly and potentially distortionary, policymakers often seek to target worker types, industries, firm types, or locations where job loss is expected to be particularly harmful. Efficient targeting, however, requires systematic evidence on when workers suffer the largest economic losses if jobs are destroyed.

This paper provides a unified and policy-oriented empirical assessment of heterogeneous earnings losses following establishment closures. We use exceptionally rich Swedish administrative data with detailed information on workers, establishments, industries, and local labor markets at the time of displacement. Closure events allow us to study well-defined negative shocks that are unrelated to individual quitting decisions and that are observed even for workers who transition directly to new employment. Following the literature pioneered by \cite{JLS1993}, we compare displaced workers to a matched control group, but embed the analysis in a flexible machine learning framework that allows treatment effects to vary jointly across a high-dimensional set of policy-relevant characteristics. Rather than evaluating one potential source of heterogeneity at a time, we show how multiple worker, firm, industry, and market characteristics interact to predict the magnitude of earnings losses. Further, we assess how well feasible targeting rules approximate our richest multidimensional benchmark.

A large and growing literature documents substantial heterogeneity in the earnings effects of job loss. On the worker side, differences have been shown to vary with age and gender \citep{DavisWachter, ichino07, hu11}, education and skills \citep{Seim2019}, occupation \citep{schwerdt2010}, immigration status \citep{bratsberg18}, family linkages \citep{Halla20}, and life-cycle patterns \citep{Salvanes2023}. On the demand side, studies emphasize task content \citep{bliendauthroth2021, yakymovych2022}, occupation-specific human capital \citep{Huckfeldt22, Braxton23}, match quality \citep{Mas2020}, sector and industry affiliation \citep{Eliason2014a, Helm22}, firm wage premia and rent-sharing \citep{Bertheau23, gulyas2025understanding, Schmieder23}, industry-level wage and union characteristics \citep{rose23}, displacement event size \citep{gathmann2020, Cederlof2020}, regional structural change \citep{arntz2022}, and aggregate conditions \citep{Eliason2006, Farber11, DavisWachter, Schmieder23}.\footnote{During 2000--2022, 25 NBER and IZA papers reported displacement effect estimates. Most estimate heterogeneous effects, with age and gender as the two most common dimensions. Related work includes \cite{reed13} on industry reallocation triggered by environmental policies, \cite{Britto22} who use GRF to study crime following job loss, and \cite{Mueller23} on predictors of long-term unemployment.}  Jointly, these studies show that displacement effects are far from uniform. While the literature convincingly documents heterogeneity along many dimensions, most of these studies evaluate one potential mechanism at a time. As a result, it remains unclear how these dimensions interact, whether they reflect distinct mechanisms or overlapping processes, and how policymakers should prioritize among them. Our aim is to integrate these strands into one unified empirical framework and evaluate their joint importance from a policy-targeting perspective.

As conventional in the displacement literature, we compare displaced workers to a matched control group. However, we refine this approach by matching on a substantially richer set of pre-displacement characteristics than in previous studies and by focusing on unconditional outcomes in order to preserve the initial balance between treated and control units. We also explore alternative specifications to illustrate the robustness of our findings. On average, job loss leads to large and persistent earnings reductions. In the short run, annual earnings fall by 24 percent and employment declines by 15 percentage points; roughly one third of the earnings loss remains ten years after displacement.\footnote{These estimates are consistent with earlier Swedish evidence, see \cite{Eliason2006} for men in the private sector and \cite{Eliason2014a} for women in the public sector.} The average effect masks substantial heterogeneity. We show that more than 30 distinct worker-, firm-, and market-level variables have statistically significant interactions with displacement when examined one at a time. 

To analyze this high-dimensional heterogeneity within a single model, we employ the \textit{Generalized Random Forest} (GRF) approach of \cite{athey2019}. GRF estimates conditional average treatment effects (CATEs) as flexible, nonparametric functions of observed characteristics by combining many causal trees. To limit overfitting, we implement cross-fitting: the data are partitioned into folds, models are trained leaving out one fold at a time, and predictions are then generated for the left-out fold. We then rank workers within each fold according to their predicted CATEs and estimate average treatment effects separately for each quantile. The dispersion is large and economically significant. The decile of workers predicted to experience the largest short-run losses see earnings fall by almost 50 percent of their pre-displacement level — 2.5 times the median loss and roughly eight times the loss in the least affected decile. Although the model is trained on short-run outcomes, the ranking also predicts long-run heterogeneity and performs well when applied to displacement events occurring after the training period. 

The results clearly show that the underlying heterogeneity is layered and multidimensional. No single observable dimension captures the underlying dispersion. Some determinants operate at the worker level, others reflect firm-specific rents lost upon displacement, and still others arise from variation in market-level re-employment prospects. This complexity complicates simple narratives and makes policy targeting more demanding than is suggested by studies focusing on single factors. 

The complexity is illustrated by age and schooling, two important features. \textit{Older} workers have much more negative effects for \textit{every level of schooling} while \textit{less educated workers} have larger effects \textit{at every age}. The magnitudes of the differences amount to 10-20 percentage points in each dimension. But, strikingly, we also find substantial heterogeneity within \textit{combinations} of age and schooling. Within nearly all 80 cells defined by categories of age and schooling, in held-out data the GRF estimates accurately identify a quarter of observations who have earnings losses above 30 percent and a quarter of observations who have earnings loss below 15 percent. Much of this heterogeneity arises from establishment or market-level factors. 

Indeed, some recent studies attribute heterogeneous wage effects to the loss of firm-specific rents at high-wage establishments.\footnote{See \cite{Schmieder23, Bertheau23, Helm22, gulyas2025understanding}. Diverging interpretations are offered by \cite{Mas2020}, emphasizing match quality, and \cite{Braxton23}, highlighting technology and occupational switching.} We add two important nuances. First, we document substantial earnings-effect heterogeneity \textit{within} closure events. Ranking displaced workers by predicted CATEs within each establishment—using only information from other events—we find that average short-run earnings losses range from 40 percent in the most vulnerable within-establishment decile to 12 percent in the least vulnerable. This implies that shared firm-level wage premia alone cannot account for the dispersion in displacement losses. Note that these results are based on \textit{earnings} rather than wages. Earnings effects are directly policy relevant (at least as a tax base) but also less demanding to identify---wages can only be observed for re-employed workers which compromises the balance between displaced workers and matched controls. 

Second, we show that a significant share of establishment-level heterogeneity can be predicted using industry and local market characteristics, suggesting that displacement costs reflect not only establishment-specific rents but also broader industry and regional conditions. This distinction is central for policy design, as policymakers may prefer place-based or industry-based interventions over policies tailored to individual establishments.\footnote{Locations with large losses (net of worker characteristics) are primarily low-density areas with high unemployment and less diversified industry structures. Industries with large losses are concentrated in manufacturing, although characteristics associated with manufacturing—such as high wage premia, low dynamism, and negative employment trends—also predict large effects outside manufacturing.}

Part of the earnings losses reflect transitions into lower-paying jobs, consistent with the findings of \cite{gulyas2025understanding}, who use GRF to show that wage losses following mass layoffs largely stem from lost firm-level rents. Our results for relative earnings highlight that this mechanism is primarily at play among workers in the middle of the effect size distribution. Among the most severely affected workers, earnings losses are instead driven by substantial and persistent employment declines, both in high- and low-paying jobs. This distinction is central from a policy perspective. Policymakers may be more concerned about non-employment than wage reductions among the re-employed, and our results indicate that the largest losses arise from the employment margin. The most vulnerable decile not only experiences the largest post-displacement losses but also exhibits a higher baseline risk of job loss and lower counterfactual earnings trajectories. One striking finding is an absence of meaningful geographic mobility toward new employment opportunities. In addition, the least resilient workers are more likely to become unemployed than others, but are not more likely to switch industries. The employment effects are both immediate and long-lasting, suggesting a potential role for early or preventive policy interventions.

Policymakers spend substantial resources on programs aimed at mitigating the consequences of structural change or transitory shocks. Examples include support schemes for workers at risk of displacement, such as the core programs of the European Social Fund (€24 billion annually), the recent Swedish “Student Financing for Transition and Retraining” program, or the many firm-support systems launched to protect jobs during the COVID-19 crisis. When designing the eligibility rules for these programs --- at least at launch, before evaluations --- it may be useful for policymakers to try to reach those at risk of large earnings losses if their job disappears. When designing targeted policies, policymakers must select which characteristics to use in targeting. Candidate dimensions include demographics (e.g. older, less educated, immigrants,..), industries and occupations (declining, non-dynamic...), or place-based indicators (rural, high-unemployment areas,..). 

To evaluate which dimensions are most informative for identifying workers who would experience large earnings losses if displaced, we compare feasible targeting rules to the multidimensional benchmark implied by the full GRF model. Our results suggest that effective targeting requires combining multiple observable dimensions. Policies based on single (but important) indicators such as age, schooling, manufacturing, or population density, capture only a fraction of the underlying heterogeneity. Allowing targeting rules to depend jointly on two observable characteristics substantially improves performance, as we show using the optimal policy tree algorithm of \cite{athey2021learning}. The algorithm shows that the best-performing two-dimensional rule assigns priority to older workers in routine occupations, effectively also capturing low-educated workers in manufacturing. As a final illustration, we use the GRF estimates to assess how the existing redistribution system insures workers with high predicted losses. Although the most affected groups are relatively better insured than other displaced workers, this pattern has weakened over time, leading to a greater pass-through of earnings losses into disposable income among the most vulnerable.

Taken together, our results show that the economic consequences of job loss are far from uniform and cannot be reduced to a single observable dimension. Some workers—young, well-educated, urban, and outside manufacturing—experience only modest earnings losses, suggesting that lost job-specific rents can often be recovered quickly. Others, even within the same closure events, face large and persistent losses driven by a combination of demographic, occupational, and market factors. This layered heterogeneity makes targeting both complex and consequential. Adverse market conditions amplify losses for already vulnerable workers, particularly through the non-employment margin. These patterns imply that carefully designed and targeted policies have the potential to mitigate both the employment and inequality costs of structural change.

The paper is structured as follows. Section \ref{S:theory} outlines a conceptual framework of heterogeneous effects. Section \ref{S:Data} describes data and identification. Section \ref{S:GRF} introduces heterogeneous effects. Section~\ref{S:adv} relates the effects to other economic outcomes. Section \ref{S:Heterogeneity} discusses the predictors. Section \ref{S:Targeting} shows targeting results. Section \ref{S:Conclusion} concludes.

\section{A Conceptual Framework}\label{S:theory}

To illustrate why the earnings effects of displacement are likely to be heterogeneous along multiple dimensions—and to motivate the identification discussion in Section 3.2—we outline a stylized conceptual framework. This framework is intended only to guide the interpretation of empirical results about heterogeneous displacement effects; it is not used for structural estimation, and we do not rely on functional form assumptions from this section in our empirical work. 

Let $D_i\in\{0,1\}$ denote displacement status. Let $J_i(d)\in\{0,1,2,\dots\}$ be worker $i$'s potential job under status $d$, where $J_i(d)=0$ denotes non-employment. Let $Y_i(d)$ denote potential gross annual labor earnings (including zeros),\footnote{Thus, implicitly we treat annual hours as part of the employment margin.} let $R_i(j)$ be annual earnings if employed in job $j>0$, and let $b_i$ be income while non-employed. Note that in practice displacement may have a direct effect on earnings beyond the job; we do not include that in this stylized model, but it would be straightforward to generalize the model. Then:
$$
Y_i(d)=R_i(J_i(d)), \qquad R_i(0)\equiv b_i.
$$

Define the individual treatment effect of displacement as $\Delta_i = Y_i(1)-Y_i(0)$.\footnote{We also report results for disposable income, incorporating taxes and transfers.} Individual treatment effects depend on observable characteristics $X$. The conditional average treatment effect (CATE) is
\begin{equation}\label{Eq:delta_rewrite}
\text{CATE}(x)\equiv \mathbb{E}\!\left[Y_i(1)-Y_i(0)\mid X_i=x\right]
=\mathbb{E}\!\left[R_i(J_i(1))-R_i(J_i(0))\mid X_i=x\right].
\end{equation}

To flesh this out further, following, e.g., \citet{Card2016}, consider a decomposition of earnings into an outside-option component and a job-specific component.\footnote{\cite{Vinay_Robin_2002}, \cite{diAddario_etal_2022}, and \cite{EK_20025} provide alternative perspectives on the separability of idiosyncratic and aggregate wage components. Although we do not rely on AKM-style wage decompositions \citep{AKM1999} in our empirical specification, see \cite{gulyas2025understanding} for an application in a related displacement context.} Then, earnings for worker $i$ in job $j>0$ can be written as
\begin{equation}\label{Eq:Rents_rewrite}
R_i(j)=\Omega_i+\beta_{ij} p_{ij},
\end{equation}
where $\Omega_i$ captures the worker's outside option, $p_{ij}$ is job-specific surplus (e.g., firm premia, match quality, or firm-specific productivity), and $\beta_{ij}\in[0,1]$ is the share of surplus accruing to the worker.\footnote{Job-specific surplus $p_{ij}$ can be interpreted as match value relative to outside options. In AKM-type frameworks, firm effects are treated as common across workers within firms. Our framework is compatible with, but does not rely on, that interpretation. Note that the functional form in (\ref{Eq:Rents_rewrite}) is not restrictive on its own, but once we interpret $\Omega_i$ as the outside option, the additivity assumption may be restrictive in combination with models of the functioning of the labor market. In principle each term in (\ref{Eq:Rents_rewrite}) could depend on $d$ (implying $R$ would also depend on $d$), but for expositional simplicity we do not include that.}  In a stylized fully competitive labor market without frictions, market power, heterogeneous amenities, or involuntary unemployment, workers' earnings would be independent of the fate of their employers. In practice, however, earnings reflect both external labor market opportunities and surplus generated within the employment relationship.

Let $q(x)=\Pr(J_i(1)>0\mid X_i=x)$, $b_0(x)=\mathbb{E}[b_i\mid X_i=x, J_i(1)=0]$, and $\Omega_0(x)=\mathbb{E}[\Omega_i\mid X_i=x, J_i(1)=0]$. For expositional simplicity, assume that absent displacement workers remain in their baseline job ($J_i(0)>0$); with displacement, workers may become non-employed or transition to a different job. Then: 
\begin{equation}
\label{Eq:CATEdecomp}
\begin{split}
\text{CATE}(x)
=&q(x)\mathbb{E}\!\left[\beta_{i,J_i(1)} p_{i,J_i(1)}-\beta_{i,J_i(0)} p_{i,J_i(0)}\mid X_i=x,J_i(1)>0\right] \\
&+(1-q(x))\left(b_0(x)-\Omega_0(x)-\mathbb{E}\!\left[\beta_{i,J_i(0)} p_{i,J_i(0)}\mid X_i=x, J_i(1)=0\right]\right).
\end{split}
\end{equation}

Heterogeneity in $\text{CATE}(x)$ thus arises from variation in (i) re-employment probability $q(x)$, (ii) the selection of individuals employed after displacement and the displacement-induced change in their distribution of jobs and match-specific earnings conditional on employment, and (iii) for those who would not be employed if displaced, the gap between non-employment earnings and the sum of outside options in employment and match-specific earnings under pre-displacement jobs. In this section we do not explicitly model worker--firm sorting (formally, what drives the distributions of $J_i(d)$); we return to this crucial issue when discussing identification below.   This paper does not attempt to separately identify the conditional expectations in (\ref{Eq:CATEdecomp}), although we present results about effects of displacement on employment ($q(x)$) and composite outcomes that relate to job switching in Section \ref{S:Mechanisms}.

Consider observable characteristics (components of $x$) that may be important for each of these sources of heterogeneity. Re-employment probabilities vary over the business cycle and across regions and industries. Further, the size of the market matters if multiple displaced workers compete on markets with few relevant jobs \citep{gathmann2020, Cederlof2020}. Both re-employment and the distribution of post-displacement jobs $J_i(1)$ may be affected by mobility costs, which in turn are related to family structure, geographic attachment, or occupation- and industry-specific human capital.  The value of non-employment depends on unemployment insurance---characterized in our setting by a capped replacement rate—--as well as family income and home production, which is related to family obligations.

Job-specific surplus will also vary for a number of interrelated reasons. Displacement is more costly when workers lose firm- or match-specific rents in settings where identical workers are paid differently across firms. In frictional labor markets with on-the-job search \citep{Vinay_Robin_2002}, workers gradually transition toward higher-paying matches and firm closures reset this process, implying larger losses for experienced and long-tenured workers. Workers with long tenure may also have accumulated firm-specific human capital through training, while wage-tenure profiles may reflect deferred compensation or tournament incentives. Downward wage rigidity may cause current pay to reflect past outside options, amplifying losses when workers are displaced from declining firms, industries, or occupations.\footnote{Specific human capital may be tied to firms, industries, occupations, or task content \citep{Huckfeldt22, Braxton23}.} Finally, although this stylized framework abstracts from fully dynamic counterfactual paths, medium-run displacement losses will be smaller in environments with high job turnover and intense reallocation where the counterfactual employment state is less stable.

The central implication of this framework is that heterogeneous displacement effects are a natural consequence of well-documented economic mechanisms such as variations in local labor demand, mobility frictions, firm-specific rents, surplus sharing, deferred compensation, and productivity dispersion across firms. Observable characteristics—such as age, schooling, tenure, industry affiliation, or local labor market conditions—may proxy for multiple underlying mechanisms simultaneously. Since effect heterogeneity may arise for very different root causes, in this paper we propose a unified empirical framework that allows the factors interact flexibly rather than evaluating them one-by-one in isolation.
\section{Data, Identification and Average Effects}\label{S:Data}

We use establishment closures to study earnings losses after displacement while estimating the counterfactual earning trajectories using matched workers from surviving establishments. This section describes data, identification, and estimated average effects of displacement. Section \ref{S:GRF} describes how we estimate the heterogeneous effects.   

\subsection{Displaced Workers and Control Workers}

We start from the Swedish \emph{RAMS} data, covering all annual transactions from establishments to employees during 1985--2017. An establishment is a physical production unit belonging to a legal entity. We use the term \say{firm} for all legal entities. RAMS is linked to other records by person, establishment, and firm identifiers. We construct an annual panel of workers aged 16 to 64, keeping only their highest-earning job.\footnote{Individuals are employed if earning at least $3$ times the 10th percentile of the wage distribution.} 

We use establishment closures during 1997--2014. This ensures 10 years of pre-displacement data and 3 post-displacement outcome years for all subjects. We define a \textbf{\emph{closure}} in year $\tau$ to occur if an establishment \textit{either} disappears \textit{or} reduce employment by at least 90 percent from $\tau-1$ to $\tau+1$. We only include establishments with at least 5 employees in $\tau-1$ and some economic activity in $\tau$. Partial layoff events are not coded as closures. We remove as \emph{false closures}, cases when 30+ percent of workers moved to a single new establishment, or to other establishments within the original firm (see, e.g., \citet{Kuhn2002}). Workers from false closures are also excluded from the control group. 

Our \emph{displaced worker} sample includes workers aged $24-60$ with $3+$ years of tenure at a closing establishment in year $\tau-1$.\footnote{Note that we include the few workers who remain within the original establishment in the cases when the establishment did not fully disappear, but where employment did decline by more than $90$ percent. Also, we do not place any restrictions on what the workers do during year baseline year. This reduces potential endogenous selection due to early leavers from declining establishments.} The age restriction ensures meaningful pre- and post-displacement measures for all workers. We use a higher upper age threshold than some previous studies since our conjecture \citep[see also][]{Salvanes2023} is that older workers suffer particularly large effects. The tenure restriction ensures that workers are sufficiently connected to the establishment, see, e.g., \cite{DavisWachter}. We impose identical restrictions on our \emph{control workers}. For closures in year $\tau$, we use control workers with identical restrictions---the one contrast being that their establishments must \textit{survive} until year $\tau+1$, with at least 10 percent of the original size. We do not impose any restrictions on individual outcomes beyond $\tau-1$. 

\subsection{Identification}\label{S:Identification}
As in the plant-closure and mass-layoff literature pioneered by \cite{JLS1993}, we estimate displacement effects by comparing earnings trajectories of displaced workers to the non-displaced control workers with similar values of $X$. We first select control workers through propensity score matching where, when estimating propensity scores, we have as predictors all variables used for the heterogeneity analysis (details below). This includes a rich set of characteristics of workers, establishments, and locations. To gain precision, we match three control workers to each displaced worker.\footnote{To estimate propensity scores, we estimate year-specific logistic regressions, and we drop individuals where overlap fails (extreme propensity scores). We match without replacement to ensure that we can split the sample into separate folds.} The identifying assumption is that displacement status $D_i$ is independent of the potential outcomes $Y_i(1)$, $Y_i(0)$ conditional on $X_i$. Under this assumption, both the average and the conditional average treatment effects --- $\text{CATE}(x)\equiv \mathbb{E}\!\left[Y_i(1)-Y_i(0)\mid X_i=x\right]$ --- are identified. The assumption requires that any \textit{selection into and out of} the closing establishments before the event that would be correlated with the potential earnings $Y_i(1)$, $Y_i(0)$ is captured by the set of observed variables $X_i$. Factors affecting the job transition process after displacement and the wage---e.g. heterogeneous bargaining power $\beta_{ij}$ and job-specific productivity $p_{ij}$ in the conceptual framework of Section \ref{S:theory}---are, however, allowed to generate selective sorting of workers to firms and/or contribute to the probability of a closure event, but \textit{only} as long as the underlying heterogeneity is captured by $X$.

Much of the earlier literature uses a similar identification approach, but studies wages rather than earnings. Wages are observed only for the employed subset, and it is well known that estimating treatment effects on wages requires much stronger assumptions (e.g. \citet{lee2009training}). Formally, in the simplified model of Section \ref{S:theory}, our focus on earnings implies that we do not need to separately identify the terms in (\ref{Eq:CATEdecomp}). Conditioning on employment induces selection on the value of a post-treatment outcome (employment) and undermines the balance achieved by matching workers with others who are similar pre-layoff. Our approach requires that conditional on pre-layoff observables, earnings-relevant worker unobservables are uncorrelated with layoffs; but it is much less plausible that the unobservables that determine being employed following a layoff are uncorrelated with wages following a layoff.  Focusing on workers who remain employed post-layoff could thus lead to spurious findings of heterogeneity, since $X$ may be correlated with the bias in estimates of displacement effects conditional on employment (e.g. when $X$ is correlated with the importance of unobserved worker quality for earnings and post-layoff employment). 

Other than the choice of outcome, our approach relies on the same identification approach as the existing literature, but we explicitly examine several potential violations of the identifying assumptions. One concern is selective worker–firm matching, whereby even after conditioning on pre-layoff observables, worker potential earnings are correlated with layoffs, e.g. higher-potential workers match with firms that are less likely to have layoffs than predicted by their observable characteristics. Moreover, violations of the assumptions in small subsets of the data (defined by regions of $X$) could be misinterpreted as heterogeneous treatment effects.\footnote{Relatedly, in finite samples subgroup differences in outcomes may vary across $X$ for idiosyncratic reasons. We therefore evaluate treatment-effect heterogeneity out of sample, reducing the risk of mistaking sampling noise for systematic heterogeneity.} 

To address these concerns, we take several steps that strengthen the credibility of our strategy relative to prior work. First, the covariate vector $X$ is substantially richer than in earlier displacement studies, capturing detailed human capital measures, employment histories, establishment characteristics, and local and industry-specific market conditions (e.g. unemployment rates, market size, and industry-level trends, cycles, and churning rates) proxying for job-offer arrival rates. We also adjust for pre-displacement earnings dynamics, which may be particularly relevant for the short-run earnings outcomes that are central to our analysis. Second, Section 4.2 presents robustness checks using alternative control groups and additional controls. We provide estimates based on comparisons of workers displaced at different points in time by exploiting future closures, thereby relying on variation in event timing, and we explore matching displaced workers to non-displaced workers at other plants within the same firm. Both approaches speak directly to concerns about selective matching on unobservables. 

\subsection{Outcomes}
We study outcomes $t$ years after displacement, from $t=-3$ to $t=10$.\footnote{All displaced and controls are observed from $t=-3$ to $t=1$. From $t=2$ on, the outcomes are missing if they have died or moved abroad. From $t=4$ onwards, information for the latest years is missing and the oldest individuals in our sample reach retirement age. Due to sampling restrictions, employment equals one in $t=-3$ through $t=-1$.} Our main outcome is annual \textit{earnings} normalized by the worker's own earnings in $t=-1$. This allows us to measure relative earnings response without conditioning our sample on future outcomes. The normalization removes fixed (over time) earnings differences across workers. We consider a binary outcome for whether workers are \textit{employed} (earn more than three times the monthly minimum wage during the year). Geographical and industry mobility is measured by whether the worker lives in a different local labor market or works in a different industry compared to $t=-1$. 

\subsection{Worker, Industry and Location Characteristics} \label{ss:characteristics}

Here, we outline our industry, location, establishment, and worker characteristics. The selection reflects variables both emphasized in prior research and available to policymakers. Details are provided in Appendix A. All characteristics are measured in year $t=-1$, unless noted otherwise. \textit{Basic demographics} include age, gender, and indicators for first- and second-generation immigrants. \textit{Family} characteristics include marital status, number of children (total and school-aged), the worker's share of household earnings, and internal migration history (born outside the current region and number of cross-location moves). These variables capture mobility costs and the value of home production.

\textit{General human capital} is years of schooling, labor market experience (years employed since $t=-10$), and annual earnings separately for the years $t=-3$ to $t=-1$. \emph{Specific human capital} is establishment and industry tenure (capped at $10$) before the closure, and variables capturing the field of education. Field specificity is measured through the share of workers that work in the ten most common industries (by field), and through separate indicators for STEM education, and for licensed fields (e.g. nurses). 

\emph{Establishment and job characteristics} are plant size in $\tau-1$, trend in plant size, average wage premium,\footnote{To avoid making the structural assumptions of the AKM model (exogenous mobility, static wage setting, and additive separability between person and firm effects), we measure the wage premium as coworkers' residual wage in the spirit of \cite{CHK2013}. The wage is residualized from age, gender and immigration status, as well as education (level and field). We measure the establishment wage premium as the deviations from the industry mean of this residual, and add the industry mean as a separate variable. For completeness, we verify that including AKM effects in addition to the information we already use does not improve our estimates of heterogeneity in post-displacement losses.} a manager dummy, routine tasks, size of the displacement event as a share of total employment in its industry-location cell, and an industry-education match indicator for workers employed in one of the 10 main industries for their educational field. \textit{Industry} characteristics, at the 3-digit NACE level, capture the industry wage premium and the industry-level dynamism and demand through churning rates, excess reallocation rates \citep{Burgess2000}, long-run employment trends, current industry-level cycles, and dummies for manufacturing and publicly funded industries. 

\textit{Local characteristics} are measured at the level of local labor markets, which are clusters of municipalities created by Statistics Sweden based on commuting patterns. Local characteristics include unemployment rates, population density, exposure to industry-specific trends, cycles, churning and reallocation rates, as well as the share of manufacturing jobs. Local industry concentration is measured by an HHI index across 3-digit industries. Furthermore, we include aggregate characteristics (year of displacement and the aggregate unemployment rate in the year $t=1$) in our analysis. 

About $180,000$ workers were displaced due to $21,000$ closure events during 1997-2014. Over $4,000,000$ workers at $200,000$ establishments are eligible as controls. Table  \ref{t:desc} in Appendix B presents sample statistics. Compared to average employees, displaced workers are more concentrated in manufacturing and fewer are from tax-funded industries. Displaced workers are also drawn from smaller establishments. As a consequence of the industry structure, men are overrepresented among the displaced. The statistics further shows that the \textit{matched} controls, as expected, are similar to the displaced. 

\subsection{Estimated Average Effects}
\label{ss:averageeffects}

\begin{figure}[H]
	\caption{Average effects of displacement}
	\label{fig1_ATE}
         \begin{centering}
      \includegraphics[width=.85\textwidth]{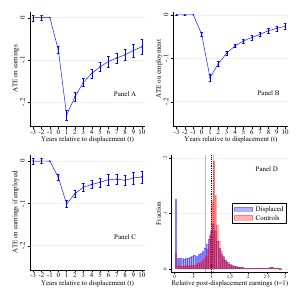}
      \end{centering}
   \footnotesize \textit{Note}: Panels A--C show estimated differences between displaced workers and matched controls (ATE) (with 95 percent confidence intervals). Panel A shows estimates for labor earnings (normalized by earnings in the year before displacement), Panel B for employment status, Panel C for labor earnings conditional on being employed. Standard errors clustered at establishment (pre-displacement) level. Panel D shows distributions of relative earnings in the year after displacement for the displaced and the matched controls. Solid lines indicate group means, the dashed line indicates unchanged nominal earnings.
  \end{figure}

We first estimate the average relative earnings effects of displacement across time. Effects are estimated as average differences in outcomes between displaced and matched controls. Figure \ref{fig1_ATE} Panel A shows familiar patterns in relation to the existing literature. The short-run impact on earnings is large, and the effects are highly persistent. Displaced workers are far from returning to the earnings and employment levels of matched controls even after 10 years. On average, earnings of displaced workers drop 24\% relative to controls by $t=1$, and are still 8\% lower in $t=10$.


Panel B shows that this earnings effect to a large extent is explained by a drop in employment, especially, in the short run. Panel C shows earnings effects for the selected employed sample, suggesting that parts of the earnings losses are driven by lower wages. However, estimates and time patterns are difficult to interpret, in particular if the effects of displacement are heterogeneous. The identifying samples change both because of the employment effects of displacement, and because employment rates by definition will fall across time also in the control group. Estimates will therefore change across time both because of time-varying imbalances between treated and controls and (if treatment effects are heterogeneous) because the underlying population changes across time.\footnote{We remove individual heterogeneity directly by using differences in earnings relative to $t=1$ instead of using individual fixed effects. This makes the treatment-control contrast more explicit as our focus is on one outcome year. But if time patterns, or treatment effects, are heterogeneous, neither of these approaches help with changes in the sample composition when conditioning the sample on employment.} The remainder of our analysis therefore focuses on earnings, which we can measure for all workers.

Figure \ref{fig1_ATE}D shows that the \textit{shape} of the earnings-change distribution (in $t = 1$) is altered by displacement, illustrating that the effects must be heterogeneous. The distribution shifts to the left and many workers drop out of the labor market entirely (zero annual earnings). Around 13 percent of the displaced (as compared to 3 percent among controls) have no labor earnings at all one year after displacement ($t=1$). In addition, there is a uniformly shaped upward shift in the number of workers with earnings above 0 but below 80 percent of pre-displacement annual earnings (the line at 1 indicates unchanged earnings). The increased number of workers with low (but non-zero) earnings show that many displaced workers work few hours during the year after displacement. The number of workers with earnings \textit{increases} above 50 percent is almost unaffected, suggesting that the number of workers who move to a better job remains unchanged.\footnote{Appendix Figure \ref{Rel_alt} presents distributions for small (to avoid spillovers) and stable (to avoid early leavers) closures, with similar results. The same figure also presents distributions for earnings changes in $t=5$ and $t=10$. For these longer horizons, differences between displaced and matched controls are smaller, but the qualitative patterns persist with a large share among the displaced having zero earnings and many workers have relative earnings below one.} 

\section{Heterogeneous Treatment Effects}\label{S:GRF}
\begin{figure}[H]
	\caption{Displacement effects interacted with worker, industry and location characteristics}
	\label{Appfig1_ATE_Het1}
         \begin{centering}
         \begin{tabular}{c}
            \includegraphics[scale=0.6]{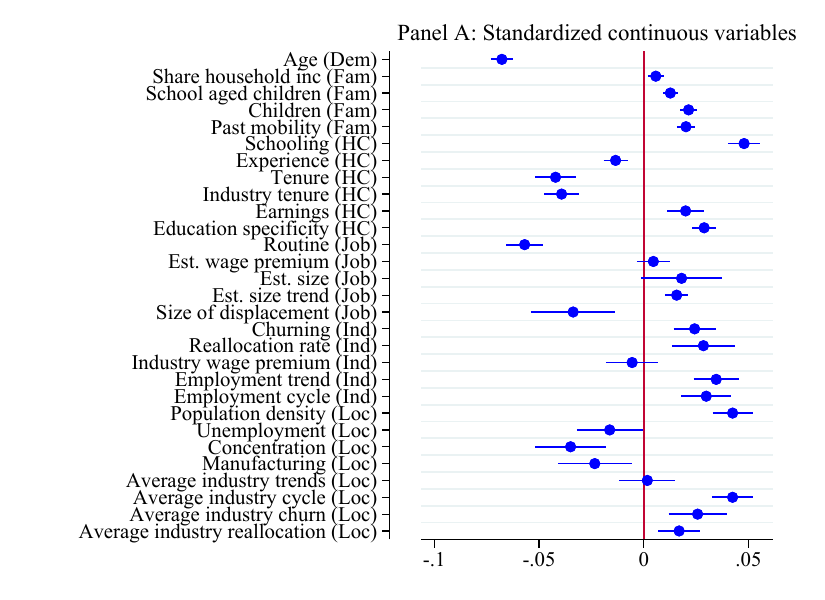}   \includegraphics[scale=0.6]{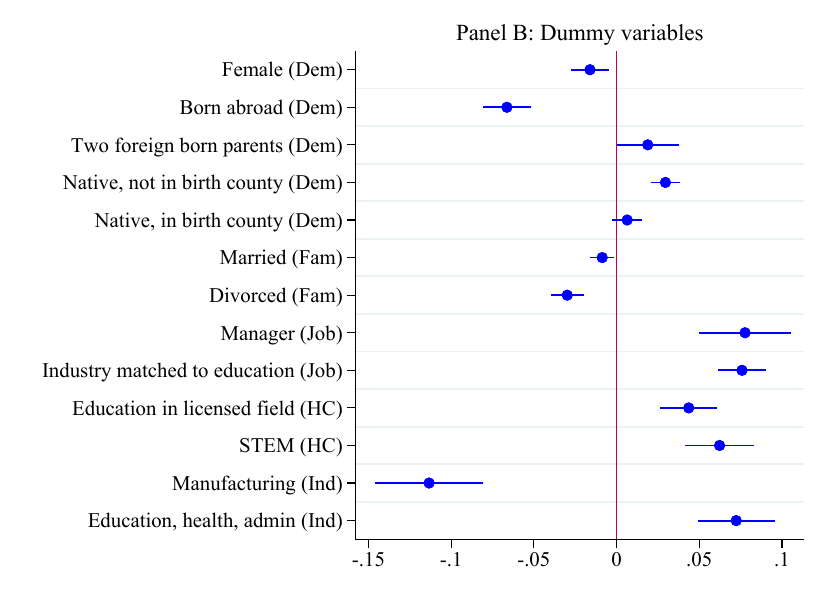} \\
         \end{tabular}
      
      \end{centering}
      
   \footnotesize \textit{Note}: Regression coefficients from separate regressions, one for each variable. Outcome is labor earnings in the year after displacement normalized by earnings in the year before displacement. Explanatory variables are an indicator for displacement, the specific variable, and the interaction of the two. The graph displays the estimated interaction terms with 95 percent confidence intervals. Standard errors are clustered at the (pre-displacement) establishment level. Panel A: Standardized continuous variables. Panel B: Dummy variables.
  \end{figure}

A conventional approach to study heterogeneous effects is to estimate linear interaction models. Figure \ref{Appfig1_ATE_Het1} reports estimates from such models where our variables, one-by-one in separate regressions, are interacted with a displacement dummy. Dummies are kept as they are, whereas all other variables are standardized to mean 0 and standard deviation 1. The estimated interaction terms are, with few exceptions, statistically and economically significant as predictors of heterogeneous displacement effects. The continuous variables with the largest estimates (in absolute values) are age (older workers lose more), schooling (educated workers lose less), routine intensity (routine workers lose more), population density (workers in urban areas lose less), tenure (longer-tenured workers lose more) and industry cycle (losses are smaller in growing industries). The dummy variables with the largest effects are the industry indicators for manufacturing (larger negative effects) and education, health and public administration (smaller effects in publicly funded industries).  All of these estimates are broadly in line with previous research  and our stylized theoretical framework. In some cases, different variables may plausibly capture the same causal processes (e.g., age and tenure), but in other cases it is more likely that the causes are different (e.g., age and publicly funded industries). 

A key challenge is how to handle so many dimensions of heterogeneity, particularly when many dimensions are highly correlated with one another. Furthermore, the underlying heterogeneity may be even more complex due to higher order interactions and non-linearities. We could gradually make the linear model richer by introducing more interaction terms and higher order interactions in the same model but as a more systematic approach to the data, we instead move to the GRF.\footnote{One alternative is to include all variables simultaneously in a simple multivariate specification that interacts the treatment with each variable. The heterogeneity patterns emerging from this approach—which ignores potential overfitting—partly differ from the GRF results reported below.} 

\subsection{The GRF}
To study multidimensional treatment-effect heterogeneity ($\mathrm{CATE}(x)$) in a flexible way, we rely on the GRF developed by \cite{athey2019}. 
 For presentation reasons, we omit the $X$ when referring to the CATE below. The GRF allows us to consider heterogeneity across a large number of covariates within a unified model while reducing the risk of overfitting. It can flexibly account for nonlinear effects and high-order interactions (e.g., between industry-specific human capital and industry trends). We use the covariates described in Section \ref{ss:characteristics}. In the main analyses, we estimate the forest using as the outcome earnings in the year after displacement ($t=1$), but below we show that our short-run CATE estimates also predict long-run effects, and further, modeling heterogeneity using as the outcome earnings in $t=3$, $t=5$ and $t=7$ gives similar results about which workers are resilient to displacement.  

The GRF iterates across random subsets of data, estimating a \textit{causal tree} \citep{athey2016} in each subset. Each causal tree is a sequence of splits. Each split partitions the data using one of the $X$ variables. The algorithm chooses the variable and cutoff value which maximize the difference in treatment effects between workers on either side of the split. In more detail, the splitting rule is based on a regression of residual earnings in $t=1$ on residual treatment assignment, where the residuals come from predictive models of outcomes and treatment that depend on the same covariates used to model heterogeneity.\footnote{See \citep[see][]{niewager2019}. We use two separate regression forests in the spirit of \cite{breiman2001} to estimate the conditional propensity score $\Hat{e}_i=W_i|x_i$ and marginal response function $\Hat{m}_i=y_i|x_i$. The treatment status and the outcome are residualized to obtain $\Tilde{W}_{i}=W_i-\Hat{e}_i$ and $\Tilde{y}_{i}=y_i-\Hat{m}_i$. The GRF is then estimated using these residualized values.} This residual-on-residual regression is used to absorb the direct impact of the covariates, avoiding that the effect heterogeneity is conflated these direct impacts. After each split, the data in the resulting nodes are split again until the workers have been grouped into \say{leaves} with similar treatment effects. 

To avoid overfitting, the tree is constructed using part of the selected subset, while treatment effects are estimated using the other part. This is one key advantage compared to simpler methods (e.g., standard regressions), in which idiosyncratic factors at the worker level may influence the estimated heterogeneity. The estimates from a single tree can be non-robust and the GRF therefore computes many trees, with each tree using a random subset of workers and a random subset of $X$ variables at each split. The complete forest is based on an ensemble of the estimated trees, ensuring that estimates are robust across subsamples, and providing consistent treatment-effect estimates. 

All steps of the estimation are \textit{clustered} at the establishment level, which is important for avoiding dependencies (leakage) across sub-samples which could lead to overfitting if outcomes are similar for workers who were displaced in the same event due to establishment-level shocks.\footnote{The clustering effectively means that all workers from one establishment are placed in the same fold, so that workers from the same establishment are not used both when building the GRF model and when estimating the CATEs. We decided to cluster at the establishment level, as establishment shocks may be particularly important for analyses of plant closures and because treatment assignment (displacement) is determined at the establishment level (see \citet{athey2019estimating, abadie2023should}.)} We set aside a test dataset containing 20\% of the closing establishments and their associated matched controls for use in the final targeting exercises in Section \ref{S:Targeting}. The remaining 80\% of the data is divided into 5 \textit{folds}, containing an equal share of closing establishments and matched controls. We then leave one fold of the data out at a time and estimate a GRF on the remaining folds. The GRF gives us estimates of the conditional average treatment effects (CATE), but we only retain estimates for workers in the \textit{left-out} fold. As a consequence, we do not use any information from the observation's own establishment closure when estimating that observation's CATE.\footnote{All of the forests we estimate contain 2000 trees. The test set used for the targeting analysis is also divided into five folds; each of the GRFs estimated using the 5-fold procedure in the training set is used to predict CATEs for one test set fold. We use the default parameters of the \texttt{grf} package in R. These are: fraction of data sampled into each tree = 0.50, number of variables randomly available for each split = $\sqrt{Total \quad number \quad of \quad x \quad variables}+20$, minimum leaf size = 5 treated and 5 control workers, fraction of sample used for determining splits = 0.5, prune leaves which end up empty when determining treatment effects = \texttt{TRUE}, maximum split imbalance = 0.05, soft imbalance penalty = 0. } 

\subsection{Estimates of Heterogeneous Effects} \label{S:Calibration}
\begin{figure}[H]
	\caption{CATE, ATE, and the risk of displacement}
	\label{fig2_CATE_ATE}
         \begin{centering}
      \includegraphics[scale=1.4]{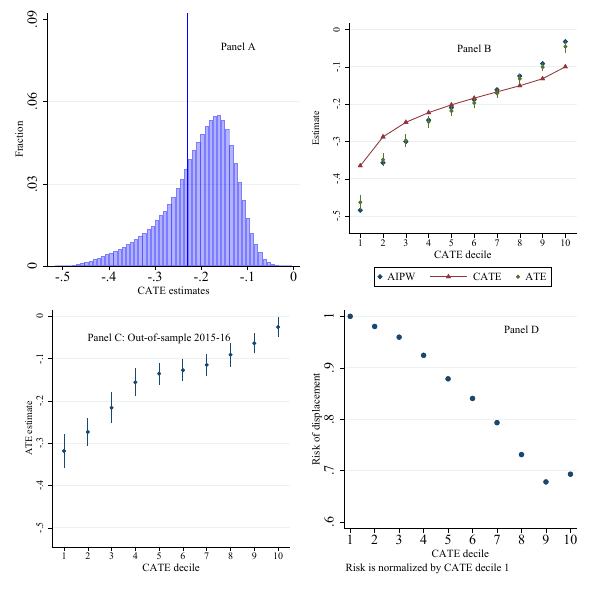}
      \end{centering}
      
   \scriptsize \textit{Note}: CATE:s for the main data set estimated using 5-folds estimation. Ranking of CATE:s for each fold separately. Panel A shows a histogram of CATE:s (line at estimated ATE for our sample). Panel B shows average CATE:s, ATE:s and AIPW scores in deciles of CATE:s for relative earnings in $t=1$. For the ATE:s, we report point estimates and 95 percent confidence intervals with standard errors clustered at the establishment level. Panel C shows ATE estimates for those displaced in the two years \textit{after} our estimation period (2015-16). We used our model to estimate CATE:s for these individuals and then divided them into deciles by CATE rank analogously to Panel B. Panel D shows the average estimated propensity of displacement by CATE decile. Scores normalized relative to CATE decile 1 to highlight relative risk. Propensity scores estimated by logit on the full sample of workers.
  \end{figure}

CATE estimates for each worker from the GRF are illustrated in 
Figure \ref{fig2_CATE_ATE}A. All CATE estimates are negative, but there is substantial variation. The CATE distribution is skewed to the right, suggesting that many workers cope reasonably well, but there is a long left tail of workers estimated to suffer very large earnings losses. 

The GRF allows us to model the CATE in a flexible and data-driven way, while guarding against overfitting towards spurious patterns. The estimated CATEs give an accurate ranking of the units in terms of effect size, but estimates are somewhat compressed to the mean, as discussed below. Thus, our preferred approach is to split the sample into CATE deciles,\footnote{For each fold of the data, the mapping from characteristics to CATEs is estimated using only data from the other folds. The decile rankings are constructed separately for each fold.} and then calculate the average effect within each decile as the difference in outcomes between treated and controls (by design, with similar values of $X$): 
\[
\mathrm{ATE}_{d} \equiv \mathbb{E}\!\left[\,Y(1)-Y(0)\mid G=d\,\right],
\qquad d\in\{1,\ldots,10\},
\]
where \(G\) indicates the decile of the predicted CATE. We refer to this as the \textit{Average Treatment Effect} (ATE), suppressing the $d$ denoting the specific decile. Under unconfoundedness it produces consistent estimates of treatment effects within the data-driven groupings produced by the GRF.\footnote{Note that, even if our CATE estimates are imperfect, our approach still provides valid estimates of treatment effects for the groups we define, so long as our unconfoundedness assumption holds.}


Figure \ref{fig2_CATE_ATE}B reports the ATEs for the ten deciles, with the largest effects in Decile 1. Reassuringly, we get monotonically more muted ATEs as we move up the CATE deciles. The differences in effects are large. Displaced workers in the least-affected decile (Decile $10$) lose only $5.9$ percent of their earnings, on average. These small effects suggest that workers in this group are able to recover (nearly) all job-specific monetary rents if they lose their current jobs. In contrast, losses for workers in Decile $1$ amount to $46$ percent, i.e. almost $8$ times larger than in Decile $10$. These estimates imply that the GRF can identify, out-of-sample, a substantial group of workers who will suffer only trivial losses, and another group who will lose almost half of their pre-displacement earnings, if their jobs disappear. 

Figure \ref{fig2_CATE_ATE}B also compares the average CATE to the ATE within each decile to assess how well-calibrated the CATE estimates are. Although the ranking is correct, the true variation across deciles (in terms of ATEs) is larger than what is predicted by the CATEs.\footnote{Formally, the best linear predictor test of \cite{chernozhukov2015} tests if the GRF can predict the correct mean effect, and variation around this mean. In our case, the estimated mean effect is well calibrated (in the full sample, the average CATE is similar to the ATE), but the variation across deciles is underestimated. For the mean we estimate $\alpha=1.06$ and for the variation, $\beta=1.57$. Both are significantly different from zero at conventional levels.} The GRF thus accurately ranks workers, but underestimates the magnitude of differences across these ranks.\footnote{One reason could be that the GRF estimates are regularized, and hence compressed toward the mean.} This pattern caries over to other cuts of the data based on CATE as analyzed below, and we therefore focus on the ATE estimates whenever we discuss magnitudes. Figure \ref{fig2_CATE_ATE}B also plots averages of doubly robust Augmented Inverse-Probability Weighted (AIPW) scores in each CATE decile, which provide alternative estimates of treatment effects.\footnote{AIPW scores are based on CATEs, as well as on the treatment and outcome models estimated by GRF. They provide a way of calibrating the CATE estimates. Alternative ways of calibrating CATE estimates are suggested by, for instance, \cite{vanderlaan2023}. We prefer to focus on within-decile ATEs for simplicity.} The small differences between the AIPW estimates and the ATE estimates suggest that our basic matching procedure is sufficient to adjust for differences in observables, and we therefore primarily report the unadjusted ATE estimates. 

\subsection{Robustness analyses}

Here we report robustness checks for internal and external validity and method choices. 

\textbf{External validity.} It is always challenging to assess the extent to which empirical results are relevant in other settings. But what we can show is that the underlying processes are stable enough to derive meaningful predictions \textit{after} our estimation period. In Figure \ref{fig2_CATE_ATE}C we apply the decile assignment to events occurring after our training data ended. Using data from $2015-16$, a full decade after the median training year, we are still able to predict treatment effect heterogeneity as illustrated by the retained monotonic relationship between CATE deciles and ATE:s. 

\textbf{Early leavers.}  Our main strategy uses matching on a rich covariate vector, and a sample selection rule that is lagged one year, to handle potential pre-closure selection out of the closing plants.\footnote{See e.g. \cite{Seim2019} on early leavers.}  To further address concerns regarding the potential that workers select out of the firm in anticipation of layoffs, we have estimated a version where we restrict the sample to closure events affecting plants with stable employment in the 3 years prior to the closure with negligible changes in results. Results are presented in Figure \ref{App:FigRobMatch}, which also reports our baseline estimates for comparison. To keep the results tractable, we focus on the displacement effect for CATE deciles 1 and 10 in these and other remaining robustness checks.

\textbf{Identification.} Our analyses rely on plant closures under the assumption that these events are independent of the potential outcomes conditional on our rich set of covariates. Figure \ref{App:FigRobMatch} reports several tests of this assumption. To examine the importance the of control variable vector, we first match on \textit{a wider set of firm variables} and adjust for an estimated plant closure-risk\footnote{Plant closure risk is estimated based on our set of industry and location variables, plant size, trend in plant size, residual wage, plant share of industry-location market, share of hires, share of leavers and firm wage quantile.} measure. The results remain robust. To test the choice of comparison group, we match current plant-closure workers to \textit{workers at plants which close in the future} (all workers who in $t=-1$ are employed at plants which close in $t=2$). The results are remarkably stable. This shows that the displacement effects we identify are not driven by plant-level unobservables. We also match our plant-closure workers to non-displaced workers at \textit{other plants within the same firm}, yielding similar estimates but substantially larger standard errors, as this restricts the sample available for the analysis. 

\textbf{Data and sampling restrictions.} In the main analysis we study establishments with at least 5 workers but we have examined alternative firm-size thresholds. Initially, to handle concerns about spillover effects from large closures,\footnote{See e.g. \cite{gathmann2020} on large closures.} we re-estimate the ATE:s in each decile using only events with less than 20 workers. We also report result when only using large firms (more than 20 workers), and when excluding workers who keep their job when plant is closed (plant closures are defined as a reduction of employment by at least 90\%.) In the main analyses, we impute routineness for parts of our sample as we do not have occupational data for the full sample. Here, we check robustness to restricting the sample to workers for which we have non-imputed occupations. In all cases, the results are virtually unchanged as reported in Figure \ref{App:FigRobMatch}. Interestingly, this also holds when splitting the sample by small and large events, suggesting that establishment size alone is a non-essential factor for the effect heterogeneity. 

\textbf{Methods.} In the main analyses, we first construct a matched sample of displaced and non-displaced using propensity score matching, and then estimate the CATEs using GRF. We have explored alternative methods which we discuss next. To examine the matching algorithm, we report results using alternative matching protocols to generate the counterfactuals, using GRF for the initial matching respectively placing more weight on aggregate variables. This leads to similar results. 

Next, we compare to CATE estimates from three different Lasso models which we estimate using a 5-folds approach to avoid overfitting. First, we run a standard Lasso, using all covariates but without squares or interactions. Second, we include squares of variables and interactions between the variables in each block of variables (e.g., demographics is one block). Finally, we implement Lasso with squares of variables and within-block interactions in a way that mirrors our GRF model (an R-learner).\footnote{First, following  \cite{niewager2019}, we residualize the outcome and the treatment using two lasso models and the observed characteristics (as described for the GRF in footnote 21). Second, we model the effect heterogeneity using Lasso by estimating a model for the residualized outcome on the residualized treatment. To avoid overfitting, all three steps are done using a 5-folds approach.} 

We use each of the Lasso models in turn to estimate a CATE for each worker. These are used to divide the sample into CATE deciles as in our main approach. The average effects of displacement in the top and bottom deciles for each classification are reported in Figure \ref{App:FigRobMatch}. Overall, the Lasso models succeed at identifying the hardest-hit workers, but are worse than GRF at identifying the most resilient workers. The simpler Lasso models are especially weak at identifying differences in the upper tail. We have also compared the MSE from the outcome and treatment models estimated using GRF and Lasso, indicating that the GRF performs somewhat better in this dimension.\footnote{The MSE of the outcome model as predicted by GRF is 0.167, compared to 0.174 for the R-learner Lasso; the MSE of the propensity model estimated by GRF is 0.180 compared to 0.188 for the R-learner Lasso. The quality of the outcome and propensity models is important for the statistical properties of resulting estimates.} Overall, we thus conclude that the GRF appears to uncover slightly more heterogeneity than a similarly constructed analysis using Lasso. 

\section{Heterogeneous Effects and Other Economic Outcomes} \label{S:adv}
In this section, we study how the GRF-classification of one-year displacement losses relate to other adverse economic outcomes. Panel D of Figure \ref{fig2_CATE_ATE} studies the relationship to the estimated \textit{risk of displacement} (using the estimated displacement propensity discussed in Section 3). The figure plots the average risk by CATE decile after normalising the risk to $1$ in CATE Decile $1$. The relationship is monotonically negative and differences across deciles are substantial; the most resilient workers (in terms of earnings CATE) have a 40 percent lower estimated risk of displacement relative to the least resilient decile. 

  \begin{figure}[H]
	\caption{Effects of displacement across time and CATE deciles}
	\label{fig3_ATE_andY1Y0}
         \begin{centering}
      \includegraphics[width=\textwidth]{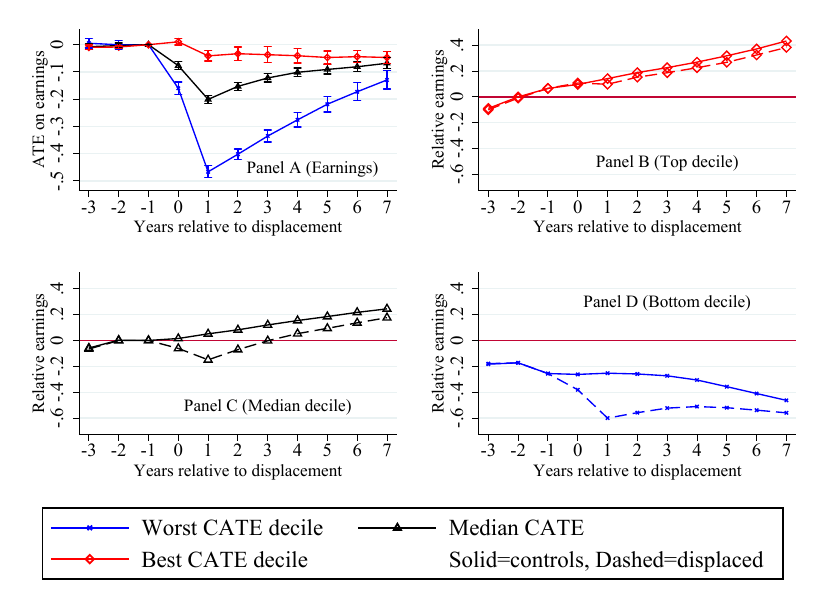}
      \end{centering}
   \footnotesize \textit{Note}: The figure shows statistics for three decile groups of the CATE distribution. The \say{median} group straddles the median (i.e. it contains the $10th$ and $11th$ ventile). Panel A shows ATE effect over time, similar to figure \ref{fig1_ATE}, but separately for the decile groups. Point estimates and 95 percent confidence interval with standard errors clustered at the establishment in $t=-1$. Panel B--D show the underlying earnings trajectories for displaced and matched controls within each decile group for the top, bottom and median deciles respectively. The series are normalized to reflect differences relative to the median group in $t=-1$. Only workers observed in each of the periods $t=-3$ to $t=7$ are included (thus we exclude 2011-2014).
  \end{figure}

Figure \ref{fig3_ATE_andY1Y0} shows how the effects of displacement evolve over time across the distribution of CATEs. We use displacement events before 2010 to ensure a balanced panel, but this does not affect our conclusions. Panel A plots ATEs over time for CATE deciles $1$ (least resilient) and $10$ (most resilient), as well as for the $10$ percent of workers straddling the median. Short-run CATEs are strongly predictive of displacement losses throughout our 7-year follow-up period, suggesting that a short-run model is sufficient to capture most of the medium-run heterogeneity. After $5$ years, earnings losses in Decile $1$ are four times larger than in Decile $10$, as compared to nine times larger in the first post-displacement year. Aggregating across the 7 years, the worst-affected decile is estimated to lose labor income corresponding to 1.98 years of pre-displacement labor earnings. This is five and a half times more than the loss of the least affected decile (0.36).

Panels B to D in Figure \ref{fig3_ATE_andY1Y0} show outcomes over time relative to the event for treated and controls, separately for the top and bottom deciles as well as the median group. As the controls provide the counterfactual outcome for the treated, the differences between these series are identical to the ATEs. Annual earnings are normalized relative to the median CATE group in $t=-1$ to highlight pre-existing earnings differences. Pre-trends are perfectly matched within each decile even though they differ across deciles. 

To make the paper tractable, we focus on CATE:s estimated for earnings in year $t=1$ and Figure \ref{fig3_ATE_andY1Y0} shows that this heterogeneity is persistent over time. To further illustrate the tight link between heterogeneous effects at different time horizons, we estimate separate GRF:s for earnings in years $t=3,5,7$ respectively, and compare the results to the baseline GRF. Appendix Figure \ref{App:Fig_grfs_t3_t5_t7} shows that we uncover similar heterogeneity across CATE deciles, regardless of which time-horizon we use to estimate the GRF model.\footnote{Specifically, the figure reports ATE estimates---for earnings in $t=3$ in Panel A and for accumulated earnings in Panel B---across CATE deciles defined by the four GRF models ($t=1,3,5,7$). The earnings losses are similar no mater which outcome horizon we use to estimate the GRF. The only model that deviates somewhat is the one trained on earnings in $t=7$, which recovers somewhat less effect heterogeneity.}

\textit{A priori}, we could have expected workers with a more positive counterfactual trajectory to suffer larger effects since they have more to lose, but the results point in the very opposite direction. The large effects for low-decile workers arise because of poorer outcomes for the treated, not because of better outcomes among controls. Workers in Decile $1$ already had lower earnings than the median before displacement \textit{and} a much less favorable counterfactual (non-displaced, control) earnings trajectory. Decile $10$, on the other hand, experiences smaller effects, but also above-median pre-displacement earnings and more positive counterfactual earnings trajectories. The figure also illustrate a key difference in terms of the catch-up over time -- much of the declining effects for Decile $1$ is related to falling earnings in the control group, which is not the case in the other deciles. Figure \ref{f:emp_potential} in the appendix replicates Figure \ref{fig3_ATE_andY1Y0} with employment as the outcome. The results are very similar, which highlights the importance of the employment margin.\footnote{A key difference when studying employment is, however, that the sample restriction ensures that everyone is employed before the event, which implies that the trends \textit{have to be negative} for all groups in Panels B to D. The declining employment rates also illustrate why we do not want to study outcomes that are measured conditional on employment, as the sample changes across treatment time. }

The results of Figures \ref{fig2_CATE_ATE} and \ref{fig3_ATE_andY1Y0} jointly suggest that job loss episodes lead to accelerated gross earnings inequality by causing larger earnings losses for workers who already had lower wages from before the event, who would have had worse earnings trajectories without displacement, and who have a higher risk of displacement. In Section \ref{S:Targeting}, we discuss the role of insurance policies that may mitigate this heterogeneity.

\section{Sources of Heterogeneous Treatment Effects}\label{S:Heterogeneity}
\begin{figure}[H]
	\caption{Differences in characteristics across CATE quartiles}
	\label{fig5_interquartile}
         \begin{centering}
      \includegraphics[width=\textwidth]{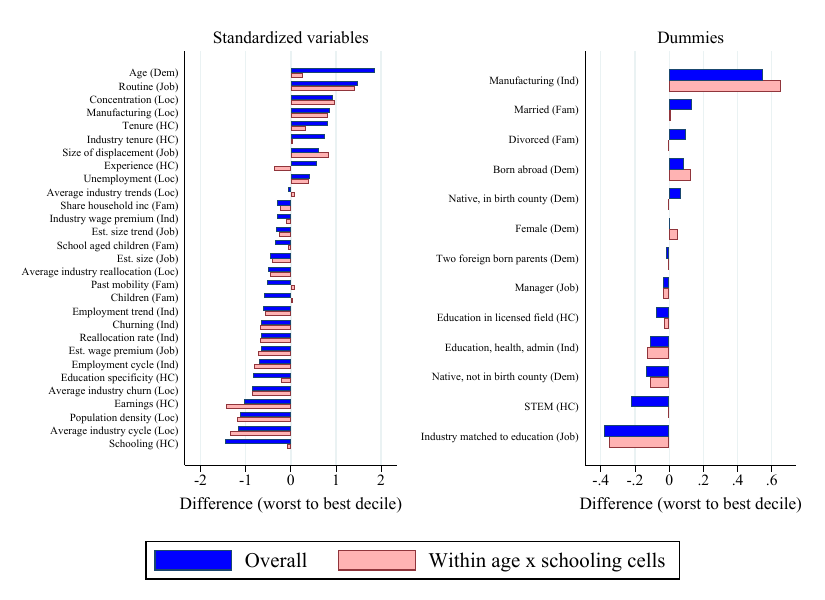}
      \end{centering}
   \footnotesize \textit{Note}: The figure shows differences in characteristics between the lowest and highest quartile of CATE:s, using the main data set. CATE:s estimated with 5-fold estimation and ranking done within each fold. The left-hand panel presents standardized (mean $0$ and standard deviation $1$) continuous variables and the right-hand panel presents dummy variables. Blue bars are for the overall quartiles and red bars cover the highest and lowest quartiles of CATE within each combination of 8 schooling and 10 age categories (see Figure \ref{fig6_contour} below). 
  \end{figure}

This section presents estimates of how the impact of displacement varies across groups of workers and market conditions, with the goal of characterizing vulnerable workers. The estimated earnings loss within each split of the data should be considered as an estimated causal effect of displacement for that worker group, but the splits themselves should not be given a causal interpretation.\footnote{For instance, when comparing displacement effects among workers with different levels of schooling, we will $i$) interpret the estimates within each education group as causal effects and $ii$) claim that the differences in estimates between the education groups describe differences in causal effects. However, we will not claim that the differences only arise because of education \textit{per se}, as education may be correlated with other important attributes, whereof some may be unobserved.}
Figure \ref{fig5_interquartile} displays differences in characteristics between workers in the lowest and highest quartiles of estimated CATE:s. Continuous variables, shown to the left, are standardized to mean zero and a standard deviation of unity. Dummies are reported at their true mean values to the right. Solid blue bars are sorted according to the magnitudes of the difference between the bottom and the top quartile. In Appendix B (Figure \ref{fig_interdecile}) we compare the top and bottom deciles instead, with similar (but starker) results.

The characteristics that stand out are the same as in the simple one-dimensional heterogeneity analysis. The quartile with the largest losses contains older workers with lower levels of schooling; these two variables have the largest cross-quartile differences among all continuous variables. Human capital variables (e.g., tenure) and occupation related factors (e.g., job routineness) also differ markedly across the CATE distribution. More vulnerable workers have less prior mobility and are less likely to hold an education in STEM fields. As already documented, vulnerable workers had lower earnings before becoming displaced. But, they also have more sector- and firm-specific human capital, as evidenced by their longer industry and job tenures. Vulnerable workers are also more concentrated in manufacturing industries and more exposed to unfavorable industry characteristics such as bad long-term industry trends and low churn rates. They are more likely to live in rural areas with high unemployment rates\footnote{This may reflect both persistent adverse local conditions and local business cycle conditions. The latter seems reasonable in light of the strong cyclicality in displacement effects observed by \cite{DavisWachter} and \cite{Schmieder23}.} and high manufacturing shares and their closure events displace a larger share of workers in their specific industry-location cell.


\subsection{Heterogeneity Across and Within Age and Years of Schooling}
\begin{figure}[H]
	\caption{Displacement effects on earnings across and within combinations of age and schooling}
	\label{fig6_contour}
         \begin{centering}
      \includegraphics[width=\textwidth]{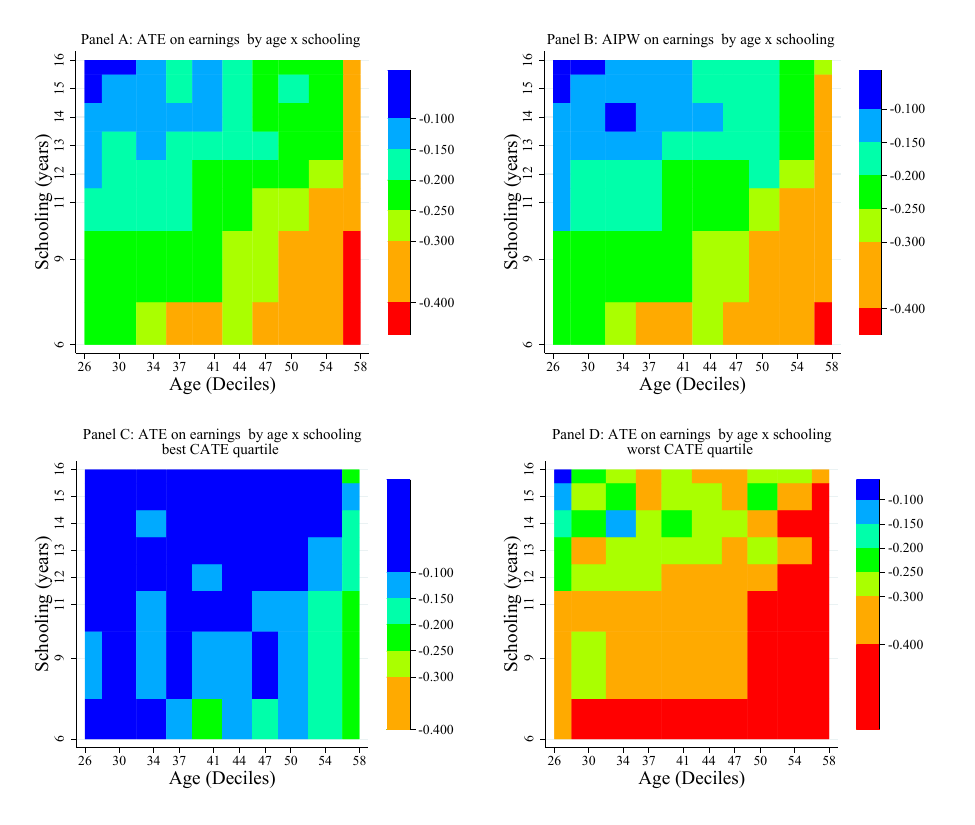}
      \end{centering}
   \footnotesize \textit{Note}: The figure divides training set workers into cells by age and schooling. Schooling has been aggregated to 8 groups by pooling the few with 10 years of schooling together with those with 9 years of schooling, and by letting the top group include all with 16 or more years of schooling. Age is defined in deciles among the displaced and the x-axis shows the median in each age group. Colors indicate the size of point estimates. Panel A shows estimated ATE:s by combinations of age and schooling. Panel B shows AIPW estimates separately within each of the 80 groups. Panel C replicates panel A, but only uses the highest quartile of 5-fold CATE:s within each group. Panel D repeats the exercise for the lowest quartile.
  \end{figure}

Motivated by the finding that age and years of schooling differ substantially across quartiles in Figure \ref{fig5_interquartile}, this section shows that older workers have larger losses due to displacement conditional on schooling, while lower-educated workers have larger losses conditional on age. Since these variables are likely to be correlated, they could, in principle, capture the same underlying factor(s). In Panel A of Figure \ref{fig6_contour}, we plot treatment-control differences (i.e. ATEs) estimated separately for $80$ combinations of age and schooling. 


As is evident from the figure, a higher age is associated with a larger ATE regardless of the level of schooling, and longer schooling is associated with a smaller ATE regardless of age. Young workers with at least a bachelor's degree lose less than ten percent of their earnings, whereas old workers who have not completed high school suffer losses of over 40 percent. Panel B shows AIPW estimates instead. These estimates adjust for any imbalances related to X-variables within each cell but, as expected, the results are almost identical to the ATEs. In the appendix (Figure \ref{Fig:AppCountourCATE}), we show that the corresponding CATE estimates are similar, but less noisy. 

In panels C and D we exploit GRF predictions \textit{within} each of these $80$ age-schooling combinations. They show ATE estimates among the most (least) resilient CATE quartile \textit{within} each combination. In (almost) all $80$ cases, the GRF identifies at least a quarter of workers with ATEs below the overall grand mean, and a quarter with ATEs above the overall grand mean. The worst-affected quartile (as identified by the GRF) of $30$ year-old university graduates has larger earnings losses than the least affected quartile of $55$-year old compulsory school graduates. 

These results illustrate that our CATE estimates accurately uncover heterogeneity (across data folds) even within narrowly defined subsets of the data. It also highlights that policy targeting is complex as job-loss effects vary both within and across worker-types. The red (shaded) bars of Figure \ref{fig5_interquartile} illustrate treatment-effect heterogeneity within age and schooling combinations. The bars show differences in means across CATE quartiles defined \textit{within} age-schooling cells. Much of this \say{within heterogeneity} is related to industry and location characteristics. A more direct illustration of the importance of industry and location characteristics is shown in Figure \ref{fig7_Extreme} in the appendix; it illustrates a strong relationship to manufacturing intensity and population density for the extreme age-education groups of those younger than 30 with at least a bachelor's degree and older than 50 with only 9 years of schooling. 
\subsection{Heterogeneity Across and Within Establishments}

\begin{figure}[H] 
\caption{Displacement effects across and within establishments}
\label{fig_within}
\begin{centering}
    \includegraphics[width=0.9\textwidth]{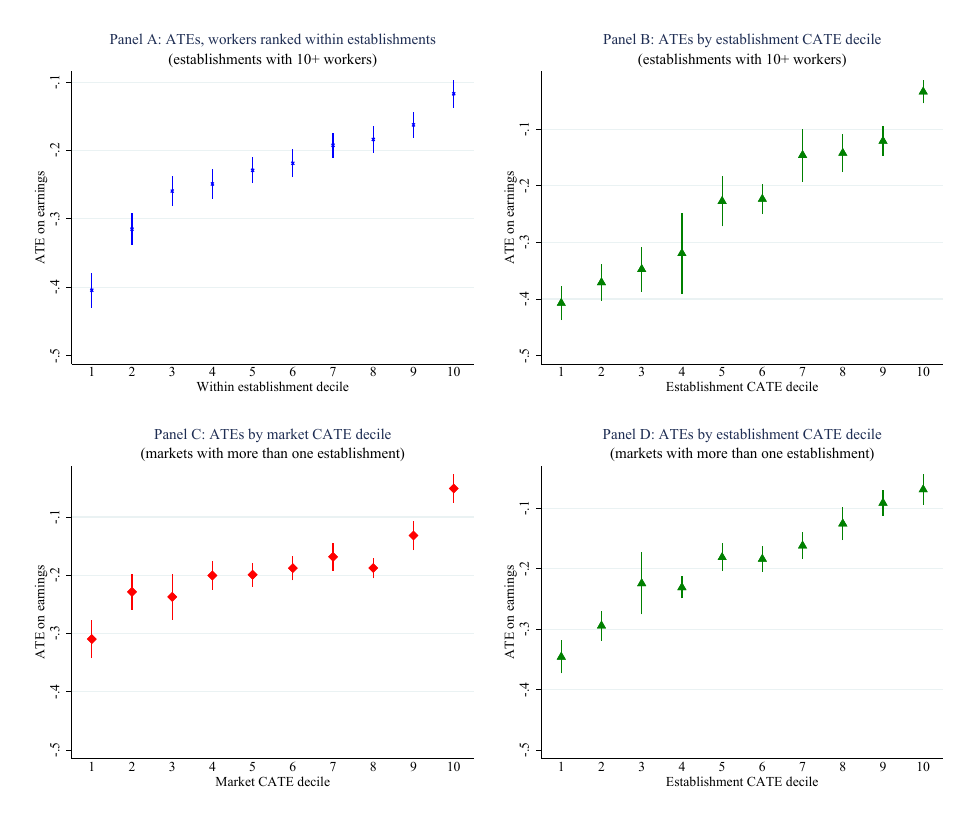}
\end{centering}

   \scriptsize \textit{Note}: Panel A shows ATE estimates when displaced workers are ranked based on their within-establishment CATE. Panel B shows ATE estimates when displaced workers are ranked based on the CATE of their co-workers (defined as the leave-out mean for the workers at the establishment and then averaging over the individuals in the CATE decile.) Panels A and B exclude establishments with fewer than 10 displaced workers. Panel C ranks the workers by the average CATE in their market (location-industry cell). Only markets with at least two establishment closures are included in panel C. Panel D ranks workers by the CATE of their co-workers like Panel B, but only includes markets with at least two establishment closures. Control workers are allocated to the same sample as the displaced worker they were matched to in all panels.  
\end{figure}

We perform a number of exercises to assess the importance of \textit{establishments} for our estimated heterogeneous earnings effects, and we find very robust evidence for heterogeneous effects both within and across establishments. To this end, we divide the sample of workers in two ways. Initially, we focus on the variation within establishments and divide the sample by their CATE decile \textit{within} each establishment. We also examine the variation \textit{across} establishments, where we for each displaced worker's establishment compute average CATE estimates (using only the CATEs of coworkers) and divide the sample of workers into deciles based on these establishment-level CATEs. We then estimate the average effect of displacement for these within-establishment deciles and across-establishment deciles. Results are shown in Figure \ref{fig_within}A (within establishment) and \ref{fig_within}B (across establishment). To ensure comparability, we focus on events with at least 10 displaced workers in both of these panels.\footnote{We need to make one further adjustment relative to the main analysis. We let the individually matched controls follow the displaced workers as we do not have estimates of CATE for all workers in the establishments of the control workers. We have verified that we get very similar estimates of average treatment effects by overall decile with this strategy, see Figure \ref{App:FigRobMatch}A in the appendix.} 

The amount of heterogeneity documented in panel A is very similar to panel B, which suggests that there is about as much heterogeneity within as across establishments. Some recent studies \citep{Schmieder23, Bertheau23, Helm22, gulyas2025understanding} point to lost rents from high-wage establishments as a key determinant of heterogeneous wage impact of job loss. Our results confirm that establishment-level factors are important, but we also find substantial heterogeneity within closure events.

Appendix B reports results from corroborating exercises. The results are identical if we use doubly robust AIPW scores (Figure \ref{Appfig_within_aipw}). The patterns are similar if we use longer term outcomes and/or if we study employment instead (Figure \ref{Appfig_within_out}). For completeness, we show (see Figure \ref{Appfig_within_AKM}) that effects are identical for high vs. low AKM-establishment effects \citep{AKM1999} \textit{within each decile}; this should not be surprising as we include establishment- and industry-specific residual wages in the GRF (i.e. when defining deciles). We further exploit the fact that CATE estimates should be correlated across coworkers from the same event if establishment-specific factors are important. Figure \ref{Appfig_within} shows that about half of the variation in CATEs across displaced individuals is shared with displaced coworkers, while the other half is specific to individuals within the same event. Finally, we show that the across-establishment ranking is more important for workers at the bottom of the within-establishment ranking (Figure \ref{fig_ATEContour}).

The predictable heterogeneity we document between workers displaced from the same establishment in Figure \ref{fig_within}A \textit{must} be attributed to individual-level factors that vary within establishments. On the other hand, CATE differences across establishments (Figure \ref{fig_within}B) are based on all variables that are shared among workers in the same establishment. Establishment-level CATEs will therefore be good predictors of treatment effects if \textit{either} establishment-level variables (the wage premium, size etc.) are important, \textit{or} if important individual/job-level variables are correlated within establishment (sorting), \textit{or} if market-level factors matter (since industry and location are fixed within event). 

In Panels C and D of Figure \ref{fig_within}, we show that the variation in displacement effects across markets (industry times location) appear to be of similar magnitude as the variation \textit{across} establishments. When characterizing markets, we use the average CATE among workers who are displaced in \textit{other} events in the same market. Results for splits by market CATE are shown in panel C. The exercise forces us to exclude singleton events, i.e. events where just one establishment is closed down in a market, and to facilitate a fair comparison we recalculate the across-establishment estimates for this sample in panel D.\footnote{Sample differences explain why results in Panels B and D are not identical.} The results show similar heterogeneity when ranking workers by the market CATEs and when using the across-establishment ranking of the workers. Note that this holds even though we rank workers by the CATEs of \textit{other} events on the same market. 

\subsection{Heterogeneity Across and Within Markets}\label{S:Aggregate}
\begin{figure}[H] 
\caption{Displacement effects across workers for different market conditions}
\begin{centering} \label{fig_within_hetero}
    \includegraphics[width=0.8\textwidth]{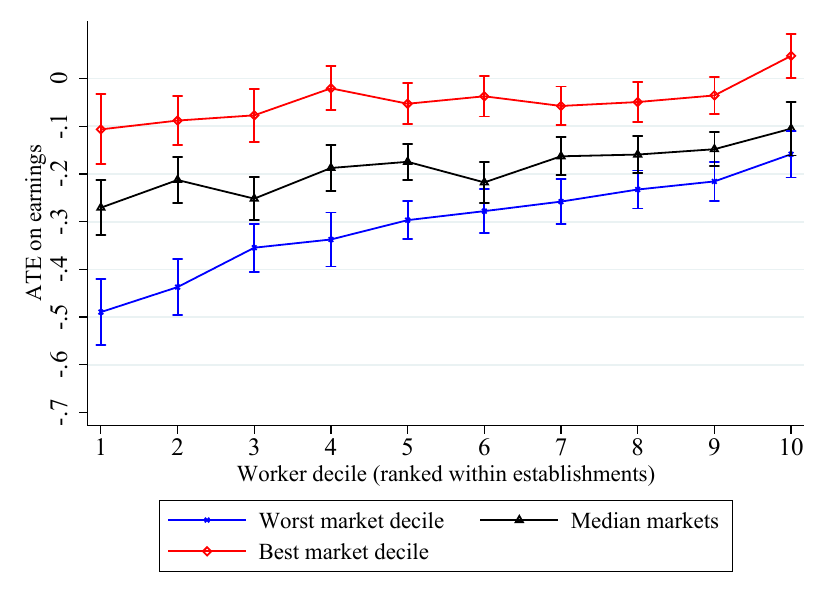}
\end{centering}

   \scriptsize \textit{Note}: The figure shows ATE estimates when displaced workers are ranked based on their within-establishment CATE for different markets, where markets are ranked by the average CATE in their market (location-industry cell). Establishments with fewer than 10 displaced workers are excluded. Control workers are allocated to the same sample as the displaced worker they were matched to in all panels.  
\end{figure}

We have seen that much of the heterogeneity within age and education combinations is related to local labor markets or industries, and that market-level factors appear to be highly predictive of displacement effects. We now show that market conditions matter more for vulnerable workers. To this end, we select the best, worst and median market deciles from Panel C of Figure \ref{fig_within}, and then examine how the displacement effects vary across workers deciles (ranking workers within establishments as in Panel A of Figure \ref{fig_within}) for these three market conditions. The results in Figure \ref{fig_within_hetero} show that vulnerable workers experience worse displacement effects under all market conditions, but the differences across worker deciles are substantially larger when market conditions are unfavorable.\footnote{Here, markets are defined based on the industry–time dimension, but we have performed robustness analyses in which markets are defined based on industry–skill–time, using different ways to define the skill dimension. This does not change the overall heterogeneity across and within markets.}


The importance of market-level factors is important from a policy perspective since policymakers are more likely to use place-based or industry-related policies than policies targeting establishments. We can also shed light on the most important industry and location variables. However, isolating the importance of market-level factors is not trivial since workers are likely to be sorted across regions and industries. For that reason we use the GRF to estimate partial dependence functions, holding the individual characteristics at their empirical levels and sequentially rotating across all observed sets of market characteristics within the same year (methodological details are in Appendix C). This way, we capture the predicted role of aggregate conditions across the full distribution of displaced workers, accounting for nonlinearities. 

The results from this partial procedure, presented in Table \ref{t: shuffle} in Appendix C, confirms that market conditions are important, and that market conditions are particularly important predictors for workers whose individual characteristics are predictive of large negative displacement effects. We also see that workers are more likely to change location if displaced under worse conditions, and workers are more likely to move away from their industry if they are displaced from an industry with large displacement effects. However, even though we find that industry mobility responds to displacement during distressful conditions, the fact that industry characteristics in general correlate so strongly with displacement effects suggests that the degree of industry mobility is insufficient to offset the negative effects of being displaced in a bad industry.

Table \ref{Tab1_indreshuffle} describes the characteristics of the best and worst locations and industries (top and bottom quartiles) after we have partialed out differences in worker characteristics across locations and industries (see Appendix C). Locations in the bottom-quartile are in particular characterized by much lower population density. These locations also have high unemployment rates and a more concentrated industry structure dominated by declining industries and manufacturing jobs. Industries with large predicted earnings losses are exclusively found in manufacturing, while industries with small predicted losses are in non-manufacturing sectors. Bad industries also have higher wage premia, are less dynamic (lower churning and reallocation rates), and experience declining employment trends over both the short and the long run. All of these attributes are typical of manufacturing, but they are also related to effect heterogeneity if we only compare different manufacturing industries to each other, or if we only compare non-manufacturing industries to each other (see Table \ref{t:manuf} in Appendix C).\footnote{Appendix C also discusses partial effects for worker attributes, net of the industry and location conditions. A main result is that the small gender differences shown in our main analysis arise because women tend to work in markets with smaller effects. Holding market conditions fixed, the GRF predicts larger effects for women.} 

\begin{table}[H]
\footnotesize
\caption{Location and industry characteristics for the worst and best locations/industries} \label{Tab1_indreshuffle}
\begin{centering}
\begin{tabularx}{\textwidth}{l zz zz}
\toprule
& (1) & (2) & (3) & (4) \\
&Worst Quartile&Best Quartile&Interquartile Difference &Standard Error of Difference\\
\midrule
\multicolumn{5}{l}{\textit{Panel A: Location characteristics, by location quartiles}}\\
Population density & 20.704 & 147.858 & -127.153 & 2.979 \\
Unemployment rate & 0.104 & 0.063 & 0.041 & 0.004 \\
Industry concentration (HHI) & 0.039 & 0.025 & 0.015 & 0.001 \\
Share manufacturing & 0.244 & 0.106 & 0.138 & 0.019 \\
Average industry trend & 0.053 & 0.112 & -0.059 & 0.005 \\
Average industry cycle & 0.004 & 0.011 & -0.007 & 0.001 \\
Average industry churn & 0.201 & 0.267 & -0.066 & 0.003 \\
Average industry reallocation & 0.143 & 0.161 & -0.018 & 0.002 \\
\addlinespace
\midrule
\multicolumn{5}{l}{\textit{Panel B: Industry characteristics, by industry quartiles}}\\
Employment trend & -0.156 & 0.268 & -0.424 & 0.084 \\
Employment cycle & -0.035 & 0.023 & -0.058 & 0.006 \\
Industry wage & 0.063 & -0.022 & 0.085 & 0.039 \\
Churning & 0.176 & 0.253 & -0.078 & 0.023 \\
Reallocation & 0.115 & 0.167 & -0.052 & 0.008 \\
Manufacturing & 1.000 & 0.000 & 1.000 & 0.000 \\
Education, health, admin & 0.000 & 0.200 & -0.200 & 0.118 \\
\addlinespace
 \bottomrule
\addlinespace
  \multicolumn{5}{p{0.98\textwidth}}{\scriptsize \textit{Note}: Panel A displays average location characteristics for the best and the worst quartiles of locations, and Panel B displays average industry characteristics for the best and the worst quartiles of industries. For details on the classification method, see Appendix C. Both panels use the partial-effects procedure described above. The location and industry characteristics are described in Section 3.2. Standard errors in Panel A clustered at the location level, and standard errors in Panel B clustered at the industry level. All estimates use the main data set.}
\end{tabularx}
\end{centering}
\end{table}

\subsection{Mechanisms and Composite Outcomes}\label{S:Mechanisms}

Our paper provides insights into the heterogeneous effects of job loss, but pinpointing the exact economic fundamentals is challenging (e.g., a full mediation analysis would require restrictive assumptions). Here, we combine our GRF output with various composite outcomes based on employment status, employment in high/low-wage jobs, and geographical and industry mobility to shed some additional light on the mechanisms. Motivated by the discussions in Sections \ref{S:theory} and \ref{S:Identification}, we analyze composite outcomes that can be measured for all subjects to avoid compromising the quasi-experimental set-up. We study heterogeneity in treatment effects on outcomes such as employment and industry transitions, but we refrain from studying wages among the re-employed (see (\ref{Eq:CATEdecomp})). For ease of interpretability, our composite outcomes are defined as categorical variables. In particular, we define new composite outcomes using mappings from outcomes related to employment, earnings, industry, and location into composite outcomes that describe mutually exclusive and exhaustive categories of the underlying outcomes.  

\begin{figure}[H]
 \caption{Mechanisms underlying heterogeneous earnings effects of job loss}
 \label{figX_mechanisms}
 \begin{centering}
     \includegraphics[scale=1.0]{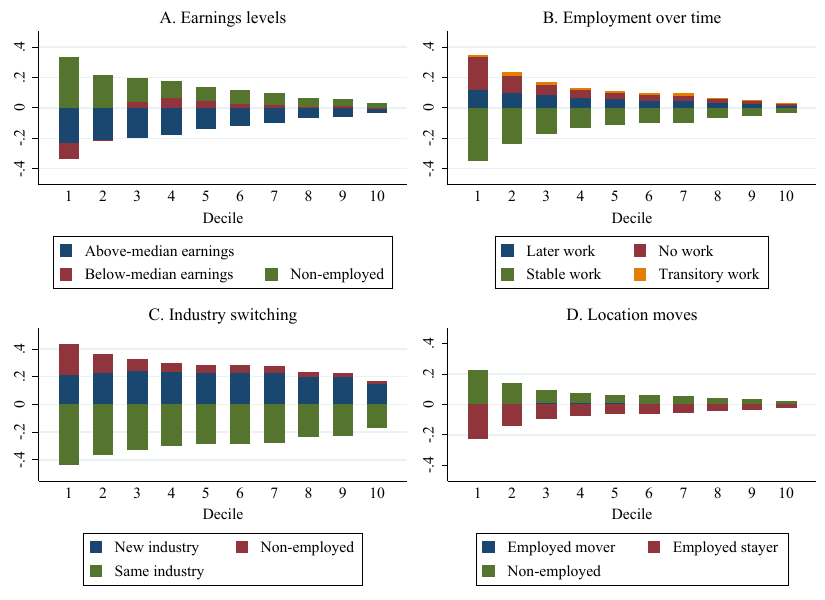}
 \end{centering}
 
  \scriptsize \textit{Note}: The figure reports average treatment effects for composite, discrete-valued outcomes within each decile of CATEs. Panel A shows the effect of displacement on the probability in $t=1$ of being employed with above- and below-median earnings, as well as of non-employment. Panel B shows the probability of being employed in both $t=1$ and $t=3$ (stable work), in $t=3$ but not $t=1$ (later work), in $t=1$ but not $t=3$ (transitory work) and in neither $t=1$ nor $t=3$ (no work). Panel C shows the probability in $t=3$ of being employed in the same or a different industry compared to $t=-1$, or of being non-employed. Panel D shows the probability in $t=3$ of being employed in the same or a different local labor market as in $t=-1$, or of being non-employed.
  \end{figure}

In Figure \ref{figX_mechanisms}, each panel shows results for a different composite measure. In all cases, the observations are grouped by deciles (illustrated on the x-axis) defined by CATEs for the outcome of earnings in $t+1$, the same groupings we introduced in Section \ref{S:Calibration}, even though we illustrate treatment effects for different outcomes (our composite outcomes) on the y-axis. Consider Panel A which shows the treatment effects of displacement on a composite outcome that takes on three values, corresponding to whether at $t=1$ the worker is employed with above-median earnings, employed with below-median earnings, or is not employed. Thus, for each decile there are three treatment effects, one for the probability of each value of the outcome. The three treatment effects must sum to zero, since the sum of the probabilities for the three values must be equal to one with and without displacement. 

Panel A illustrates that the large job-loss effects in the hardest-hit deciles reflect a shift from both high- and low-paid jobs to non-employment. In contrast, for deciles 3–6, high-paid employment declines while both non-employment and low-paid employment increase. This suggests that these workers cope somewhat better with job loss because they are partly able to transition into low-paid jobs rather than exit employment.

Panel B examines employment dynamics focusing on a categorical composite outcome based on employment in periods $t+1$ and $t+3$. Panel A shows that the effect of displacement on the “no work” value of the outcome, corresponding to no employment in neither $t+1$ or $t+3$, is much more important relative to the “later work” (employment in $t+3$ but not $t+1$ response) in the hardest-hit deciles than in the top deciles. It indicates that prolonged periods without a job contribute to explaining the effect heterogeneity at the bottom. In Panel C, we focus on a categorical outcome based on industry mobility. Industry mobility in absolute terms is roughly the same (about 20 pp) across all deciles—the more resilient deciles are just better equipped to return to the industry from which they lost their job rather than leave employment. Panel D shows that geographical mobility plays a limited role, as job loss has a negligible impact on the probability of moving to work in a new commuting zone across all deciles.

\section{Policy Targeting}\label{S:Targeting}
Given the severe average impact of displacement, policymakers have designed various forms of policies to ameliorate the consequences of job loss and/or to prevent it from occurring (e.g. through short-time work schemes or employment protection legislation). 
Several factors, such as policy heterogeneity in the employment effects and costs of the policy interventions, are relevant for policy targeting. However, prior to available evaluations that inform about such heterogeneity, policymakers may benefit from targeting workers who face a high risk of substantial earnings losses following job displacement. We therefore consider policy targeting based on earnings losses, and as shown by our analysis above, such targeting based on a single variable is unlikely to capture the complex heterogeneity pattern uncovered by the GRF. But for policy purposes we would like to find usable (i.e. simplified) targeting principles that would direct policymakers toward the right set of workers. 

As a first step, we consider key one-dimensional predictors at the worker level (age, schooling), industry level (manufacturing) and location level (population density), and pairwise combinations of these. To make further progress, we then select targeting variables using optimal policy trees \citep{athey2021learning}. We consider trees of depth two, selecting the two best targeting variables from our full set of variables. Each set of variables defines a targeting rule which we use to extract a fraction of targeted workers (varied between 5 and 25 percent of the overall sample) and estimate the displacement effects on earnings one year after displacement (ATEs) for the selected group of workers.\footnote{ Whenever there is a tie, we select a random set within this tie.} For targeting using the full GRF, this implies selecting the workers with the most negative CATEs, for age it means selecting the oldest workers, and so forth. To obtain a fair comparison between the GRF and the simple targeting rules we use our held-out test set, i.e. 20 percent of the closing establishments and their matched controls not used for producing any of the results shown so far. 

\begin{figure}[H]
 \caption{Displacement effects for workers selected by different targeting policies}
 \label{fig8_targeting}
 \begin{centering}
     \includegraphics[scale=1.0]{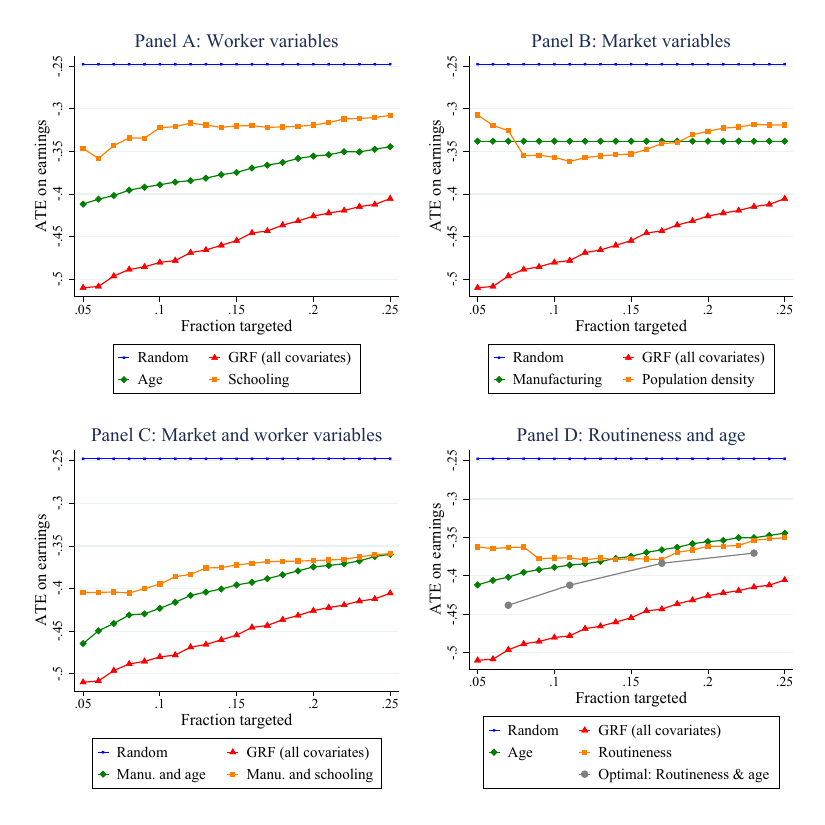}
 \end{centering}
 
  \scriptsize \textit{Note}: The figure reports estimated observed displacement effects on earnings one year after displacement (calculated as displaced-control differences, "ATEs") when selecting a fraction of the workers using different targeting rules. The random rule selects a random set of workers. The GRF selects workers using the estimated CATEs. Panel A: the age rule targets the oldest workers first and the schooling rule the least educated first. Panel B: the manufacturing rules targets a random set of manufacturing workers, and the population density selects workers in the least dense locations first. Panel C targets the oldest respectively the least educated manufacturing workers. Panel D reports targeting results when targeting the oldest workers (age) and workers in the most routine occupations. It also reports results when targeting on age and routineness using optimal policy rules \citep{athey2021learning}.
  \end{figure}
The results are reported in Figure \ref{fig8_targeting}. The blue lines in all panels show the displacement effect among a randomly selected set of workers. This results in an average earnings loss of 22.2 percent regardless of how many workers we select. On the other extreme, when using the full GRF model which uses all covariates, we find, as expected, a group of workers with much larger displacement effects, as shown by the red line. 

The five percent of workers who are hit the hardest according to the GRF experience displacement losses of 50 percent. Among the hardest-hit 25 percent, earnings losses are 40 percent. Panel A also includes targeting based on age (target the oldest first) and schooling (target the least educated first). The green line shows that average earnings losses are 43 percent among the oldest five percent of workers, and 33 percent among the oldest 25 percent. Targeting based on schooling, shown by the orange line, reaches workers who are on average less hard-hit than the oldest workers (e.g., losses of 39 percent when targeting the five percent least educated). We also see that targeting one variable (age or education) is substantially better than random, but also clearly worse than targeting based on the output of the GRF. For presentation reasons, Figure \ref{fig8_targeting} does not include standard errors, but GRF-based targeting is significantly better than random, age-based and schooling-based targeting at conventional significance levels. The same holds when GRF-based targeting is compared to other simple targeting rules.\footnote{To examine these differences in more detail, we have also explored the rank-weighted average treatment effects (RATE) metrics from \cite{yadlowsky2021}, with results for the Qini coefficient metric in Table \ref{Tab4_RATE} in Appendix B. It shows that prioritization based on the CATE:s from the full GRF performs significantly better than a random allocation, confirming the presence of heterogeneous treatment effects. Targeting based on the GRF model also outperforms targeting on age or schooling, but these simple targeting rules perform significantly better than a random allocation. We obtain similar results for all the other simple targeting policies evaluated below.}

Panel B of Figure \ref{fig8_targeting} considers targeting on industry and location characteristics. If manufacturing workers are selected at random, the losses of the workers reached by the policy are around 33 percent (green line). Since manufacturing workers constitute more than 25 percent of our sample, this average loss is the same when targeting the hardest-hit 5 percent and the hardest-hit 25 percent. However, Panel C shows that it is possible to improve targeting by selecting the oldest or least educated manufacturing workers. The first five percent of individuals selected according to these rules suffer earnings losses of 46 percent; if a quarter of the individuals in the test sample are selected, their losses amount to 36 percent. Targeting workers in locations with low population density (shown by the orange line in Panel B) does slightly less well in terms of identifying the hardest-hit individuals than targeting based on manufacturing. 

In Panel D, we explore targeting rules based on optimal policy trees \citep{athey2021learning}. Specifically, we assume that we have a fixed per-person cost of the policy and that the benefits of the policy are proportional to the income losses after displacement. We consider different costs of the policy in order to generate different fractions of targeted workers (the figure shows the preferred targeting of three optimal policy trees which assume different costs of the policy). When choosing between all of our variables, the preferred rule involves targeting based on age and the job's routine content. For a policy cost that results in about 10 percent treated, the optimal policy is to target everyone older than 59, and to target workers aged 51--58 if the level of routineness in their jobs is above the 80th overall routineness percentile (corresponding to the 69th percentile in the sample of displaced workers). Most likely, age is selected because age captures losses related to occupation-specific and establishment-specific human capital, as well as lost job-ladder related rents in the pre-displacement job. Routineness may capture exposure to both declining demand (e.g., related to the manufacturing industry) and low education. The gray line in Panel D shows that using information on both routineness and age improves targeting compared to targeting based on age or routineness only, but still performs worse than targeting based on the entire GRF. 

We conclude that simple targeting policies give better results than randomization, but do not quite reach the performance of the GRF model. This confirms our key finding that the size of displacement losses cannot be explained by any single variable in isolation as it takes at least two variables to come close to the GRF. The GRF's ability to take many characteristics into account in a flexible way is key in this context. If GRF estimation based on all characteristics in our analysis is unfeasible, a targeting rule that combines information on routineness and age works best.

\subsection{Targeting and Existing Redistribution Policies}\label{ss:insurance}
Public tax and transfer systems are designed to mitigate the pass-through from gross labor earnings to after-tax income by means of, e.g., unemployment insurance and social assistance, and progressive taxation. The degree of insurance therefore tends to be larger for workers who experience large gross earnings losses. In addition, as outlined in our theoretical framework, workers could experience larger earnings losses because they are well-insured in case of unemployment. Both of these channels suggest that the degree of insurance should be positively correlated with the size of the earnings effect. To illustrate this, we examine the impact on disposable income by labor-earnings CATE decile in Appendix Figure \ref{fig4_dispinc}.\footnote{Disposable income is calculated by Statistics Sweden at the household level as some support systems give money to households, not individuals. The income is then attributed to household members according to a fixed formula. Thus, the variable also captures within-household income pooling with other household members, but our results are very similar if we focus on singles.} As expected, disposable income losses are much smaller than labor income losses, in particular in Decile $1$ where the degree of implied insurance is $71$ percent (as compared to around $50$ percent in Deciles $6$ to $10$).\footnote{Adding taxes to our theoretical framework, we can write the disposable income effect as \newline 
$\Tilde{\Delta}=(1-q)b +qE((1-tax_k)W_k) - (1-tax_j)W_j$ 
and the degree of insurance as 
$ (\Delta-\Tilde{\Delta})/\Delta $ where $\Delta$ is the treatment effect on gross earnings.
} Thus, insurance responses mitigate the impact on inequality, but instead shift parts of the financial burden from the most affected workers onto public finances. 

Because insurance policies tend to focus on responses at the bottom, it is not obvious that the groups that suffer from the largest gross earnings losses always lose more in terms of disposable income. To study this issue, we estimate a separate GRF for disposable income. The correlation between CATEs for the two outcomes is positive, but moderate ($\rho=0.37$). In Appendix Figure \ref{fig4_dispinc2} we show the interquartile ranges of characteristics of workers as a function of disposable income losses. Overall, large disposable income losses are relatively more prevalent higher up in the income distribution. Using data from Canada, \cite{stepner} also finds that tax and transfers systems provides considerable insurance against economic shocks (layoffs and illness), and that this insurance is relatively less important at the top of the income distribution. Two contributing factors are social transfers and progressive taxes. 

  \begin{figure}[tbh]
	\caption{Displacement effects on earnings and disposable income over time and across earnings CATE deciles}
	\label{fig_time_dispinc}
         \begin{centering}
      \includegraphics[width=\textwidth]{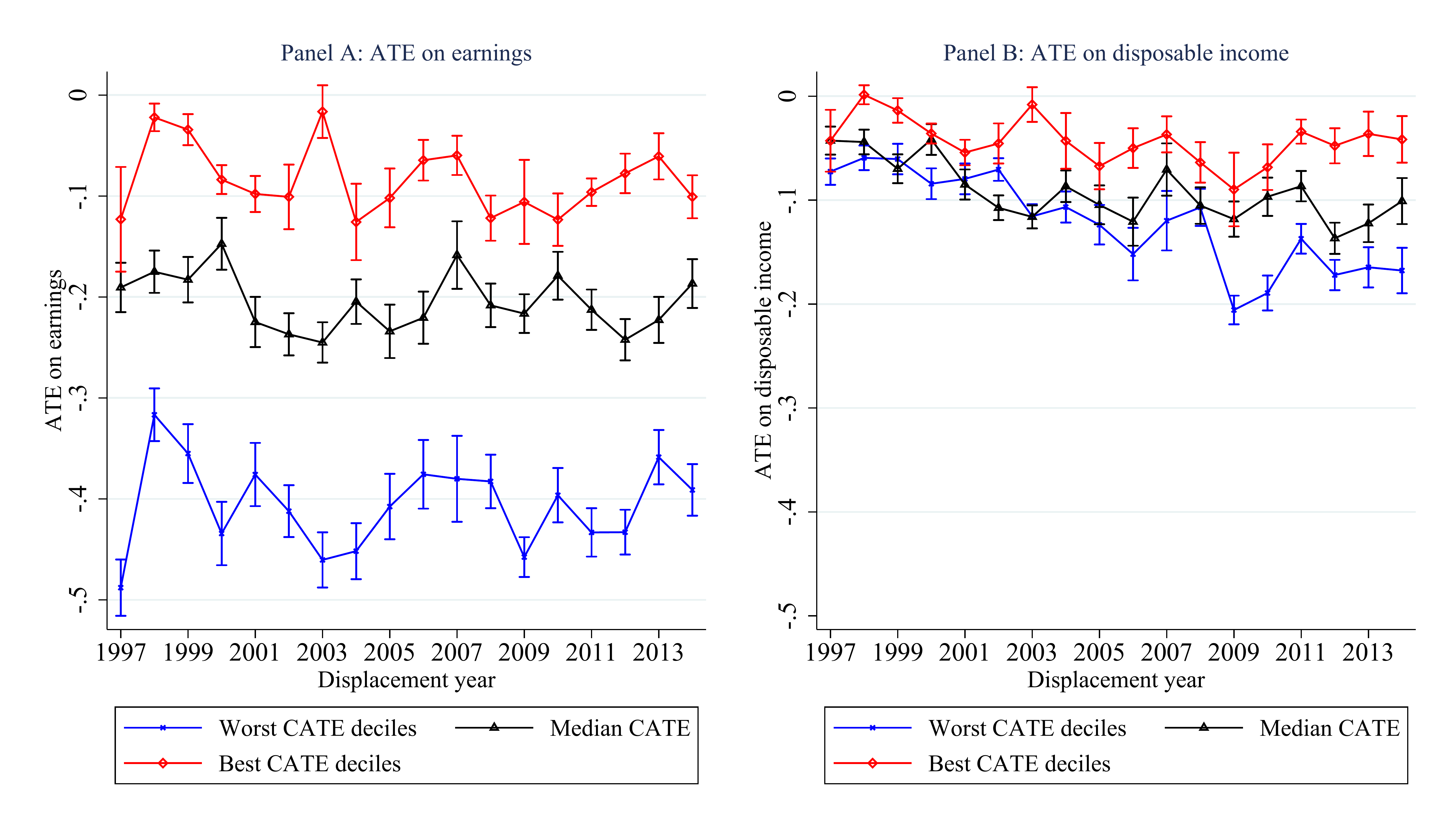}
      \end{centering}
   \footnotesize \textit{Note}: The figure shows estimates by displacement year and deciles of the CATE on earnings in $t+1$. CATE deciles defined within years. Panel A shows ATE estimates on labor earnings in $t+1$ and Panel B ATE estimates for disposable income in $t+1$. Worst CATE deciles includes the first and second CATE deciles, Median CATE the fifth and sixth CATE deciles, and the Best CATE deciles include the ninth end tenth CATE decile.  
  \end{figure}

An interesting evolution, which our data allow us to shed new light on, is the gradual erosion of the generosity within the Swedish Public Insurance System \citep{KR2017}. The generosity declined as benefit caps and benefit floors remained fixed in nominal terms over most of our sample period, while nominal (and real) wages grew considerably. In Figure \ref{fig_time_dispinc}, we show how the impact of job loss on labor earnings and disposable income evolved over time across the distribution of predicted earnings effects (CATEs for earnings in $t+1$). The effects on labor earnings are relatively constant across the period, but the patterns related to effects on disposable income diverge very strongly. As is evident from the graph, the pass-through to disposable income has increased substantially in the group with the largest predicted earnings losses. This implies that the gradual policy changes that occurred during the study period eroded the degree of insurance particularly among workers who experience large negative effects on gross earnings when displaced. 
 
\section{Conclusions}\label{S:Conclusion}
The results presented in this paper contribute to the vast displacement literature by illustrating how heterogeneous earnings effects of job loss can be estimated, and how this may help policy makers or caseworkers if they wish to target workers facing particularly adverse effects of job loss. Our results show that groups with large negative effects are vulnerable in other economic dimensions as well: they have lower pre-displacement earnings, a higher displacement risk, and much worse counterfactual earnings trajectories than other displaced workers. This suggests that job loss leads to accelerated gross earnings inequality, which may motivate well-targeted policy interventions.

The existing literature discusses many explanations for why post-displacement losses vary, highlighting factors such as firm, industry and occupation-specific human capital, location mobility and aggregate conditions. Our results confirm that determinants of earnings losses after displacement are multidimensional, as substantial systematic heterogeneity remains after conditioning on crucial predictors such as age and schooling, or on the closing establishment. While older workers with lower levels of general human capital lose more on average, even old and low-educated workers can do better than the average worker if other circumstances are favorable. Aggregate demand-side conditions at the local and industry level play an important role as workers in urban areas and outside of manufacturing consistently fare better than other displaced workers. 

The complexity of heterogeneous effects makes policy targeting challenging, but also implies that policy-design details may affect whether or not a policy reaches the intended target group. Our results show that targeting on age, education, manufacturing indicators or population density is a substantial improvement over a random allocation if the aim is to target workers with large effects. But the overall heterogeneity is too complex to be  fully approximated by one-dimensional targeting rules. The targeting rule which comes closest to the targeting recommended by the GRF model involves focusing on older workers in routine occupations, which is a group which suffers from both detrimental individual and market-level conditions. 

In general, our results indicate that the economic value of jobs is highly heterogeneous. Some workers, primarily the young, well-educated and those in non-manufacturing industries, are able to move on to a new job with only short-term and marginal losses if their establishment closes down. This pattern is consistent with a labor market without unemployment and where any job-specific wage rents can be fully recovered if the current jobs disappear. Other groups of workers, such as the older and less educated, instead suffer very large and prolonged losses if their jobs disappear, suggesting that these workers are exposed to highly imperfect labor markets. The interaction between worker and market characteristics should also be understood in this context. Workers with favorable individual characteristics experience small effects of job loss regardless of market conditions, whereas workers with large average effects are much more sensitive.

Our analysis also points to a number of interesting avenues for future research where predictions from the GRF model can help us understand other aspects of the economic and social consequences of job loss. The predictions can be used to analyse a wider set of labor market outcomes beyond earnings, which have been at the focus of this paper. Furthermore, the predicted heterogeneity is likely to be useful when analysing different back-to-work pathways such as further education or self-employment that members of different groups may rely on. Most notably, the GRF model can be used to scrutinize the targeting embedded in existing support policies, as illustrated by our analysis of redistribution policies in the Swedish context. This approach can be extended to assess how policy targeting varies across specific policies such as benefit schemes, short-time work schemes and life-long learning policies.   

\bibliography{References}
\clearpage
\appendix
\counterwithin{figure}{section}
\counterwithin{table}{section}

\section{Online Appendix: Data Details}

Below, we explain the variables used in the GRF model and when matching displaced workers to non-displaced controls. 

\subsection{Demographics}

Data on basic demographics are drawn from the administrative population register Louise produced by Statistics Sweden. Data in Louise is collected in November of each year. \textit{Age} is measured in years. We include a \textit{Female} dummy. A categorical \textit{Native-immigrant} variable takes the value zero for natives, one for those with two foreign-born parents, and two for those born abroad (unless both parents are born in Sweden).

\subsection{Family} 
Civil status is coded using two dummies. $Married$ workers are either formally married, or cohabit with a partner with whom they have a common child. This definition is well in line with Swedish perceptions of "marriages" and is used in most research on Swedish data. $Divorced$ is a dummy for workers whose marriage has ended and who are not currently defined as married according to $Married$. 

To quantify the economic value of the partnership and degree of dependence, our \textit{Share of household income} variable measures the subject's labor earnings as a share of total household labor earnings. 

We measure the number of children in the household using two different variables. \textit{Children} contains the number of all children under the age of 18 in the household, whereas \textit{School-aged children} counts children aged 7 to 17. We make this distinction to allow the second variable to measure the additional impediments to mobility that may arise when children start school at age 7.

Social ties to the current location are measured by two variables. First, we define a dummy for being \textit{Born outside of the current county}. This variable naturally takes the value 1 for all foreign-born.\footnote{The register information is based on the county of birth and cannot be disaggregated further.} We also measure \textit{Past mobility} as the number of moves across local labor market boundaries during the past 10 years. This variable is a lagged counterpart of our key indicator for geographical mobility after the event.

\subsection{General Human Capital}
\textit{Years of schooling} are based on the highest achieved level of education. \textit{Labor market experience} is measured as the number of years employed during the 10 years prior to displacement. Because of the tenure restriction, all displaced and control workers must have been employed for at least three years prior to the year $t=0$ and the variable therefore has a range from 3 to 10. 

In this category, we also include pre-displacement earnings. We measure these by \textit{Earnings} as the rank among the full population of displaced and controls in year $t=-1$, as well as two variables capturing \textit{Earnings growth} from $t=-3$ and $t=-2$ respectively to $t=-1$. The rank form of the \textit{Earnings} variable has been chosen to ensure orthogonality to time trends.

\subsection{Specific Human Capital}
It is likely that workers differ in terms of how costly or difficult it is for them to switch firm and industry if hit by a negative shock. The ability to adjust along this margin is potentially determined by the degree to which the worker is tied to the closing firm or sector. In order to measure the empirical relevance of these aspects, we include variables capturing pre-closure establishment \textit{Tenure} (truncated at 10), and the number of years spent in the same industry as the closing establishment (\textit{Industry tenure}, also truncated at 10). 

Furthermore, we characterize the \textit{Education specificity} using data on the 1-digit level and 3-digit field of the highest achieved education (thus, high school programs and college majors). Our metric uses the fraction of workers, by education cell, that is employed in the ten main 3-digit industries. The strategy follows \cite{Altonji2012}.\footnote{An alternative is provided by \cite{LeightonSpeer2020} but their approach requires a  full matrix where all types of education are present in all industries and is therefore less well suited to our granular data.} Examples of fields with high levels of specificity are pharmacists and nurses. 

We further include an indicator for $STEM$ education beyond high school. We also provide a dummy for \textit{Education in a licensed field}, that is a field that caters to the health or education sector.

\subsection{Lost Job Characteristics} 
The causal impact of job loss for displaced workers does not only depend on their outside opportunities, but also on aspects of the job that they have lost. We therefore define a number of variables which capture key characteristics of the lost job. 

We first characterize the lost job by \textit{Plant size} in terms of employment in $t=-1$, and \textit{Plant size trend} between $t=-3$ and $t=-1$, measured as the difference in log(\textit{Plant size}). 

Additionally, we measure the \textit{Plant wage} premium at the closing establishment. This feature is shown to be particularly important in the case of Austria by \cite{gulyas2025understanding}. They study mass layoffs and characterize the affected firms by firm wage effects as estimated conditional on person-fixed effects in the tradition of the AKM model \citep{AKM1999}. Indeed, there is a large literature discussing the origins of firm-specific rents, and how they should be estimated and interpreted \citep[see, e.g.,][in the context of plant closures]{card2018, BLM2019, Mas2020}. A drawback of the AKM model is that it requires structural assumptions that we do not want to impose, and is likely to result in biased premium estimates for dying firms. We therefore take a slightly different route. As a reduced form measure of the pay level of the closing firm at the time of displacement, we use the leave-one-out mean of residual earnings in year $t=-1$ as in \cite{CHK2013}. For each displaced worker and potential control, we characterize the closing establishment environment by the average Mincer residual earnings of all co-workers at the same establishment. Residual earnings are computed using year-specific regressions of log earnings on 3-digit industry indicators, years of schooling, field of education (2 digits), gender, immigration status and a full set of age dummies.\footnote{The $R^2$ in a regression of labor earnings on the plant wage premium measure among the displaced is around 22 percent.} 

We measure the \textit{Routine} task component of the lost job. To this end, we use data on routine intensity by occupation from \cite{autor2013} and \cite{goos2014}. We translate the occupational codes into the Swedish nomenclature. We have data on occupations for about half of our workers. We impute routine intensity for each worker based on the average routine score for those in the same detailed education-industry combination.\footnote{We define the combinations using 1-digit education level, 2-digit education field, and three-digit industry. For cells where we have fewer than 100 workers we use data on the 1-digit industry and education (field+level) instead. The leave-out correlation between our education-industry-based measure of routineness and routineness as measured directly based on occupation is 0.7. The reason for not using occupation-level routineness directly is the large number of missing values this would imply.}$^,$\footnote{Robustness analyses indicate that our results are robust to imputing routine intensity for half of the sample. Our results are robust using only data where information on occupations is available. Moreover, broad occupational categories do not appear to predict displacement effects beyond job routineness and the other covariates, suggesting that detailed occupational data for all workers may not be crucial for predicting heterogeneous job-loss effects.} 

We further construct a \textit{Manager} dummy for workers who are employed as managers. This information is drawn from the occupational codes, which we do not have for all workers, and therefore contains false negatives. To mitigate the problem, we impute manager status based on data from the previous three years if the information is missing in $t=-1$. 

We calculate the \textit{Relative size of the displacement event} as the share of total employment by 3-digit industry and local labor market combination. This is motivated by previous studies, e.g. \cite{Cederlof2020} and \cite{gathmann2020}, that have found that the impact of being displaced in a large event is particularly severe. One possible explanation for this finding is that workers from the same event may compete with each other for the same job openings, and this type of competition might be particularly problematic if the displacement event is large relative to the industry-specific local labor market. 

Finally, we generate a variable for \textit{Industry being matched to education}. This indicator takes on the value 1 for workers who were employed in one of the 10 main 3-digit industries for their 1-digit level and 3-digit field before displacement. 

\subsection{Industry Characteristics} 
There is ample evidence that, on average, workers have comparative advantage in their industry of employment and that shocks to this industry have an impact on their overall earnings prospects. Examples include \cite{Carlsson2016} for Sweden, who show that that technology shocks have a larger impact on workers' wages if the shocks are shared with other firms in the same industry and that this distinction is entirely driven by workers with fields of education where most job-to-job mobility is within the industry. Similar arguments are made in \cite{LMS}. It is also likely that displacement has more lasting negative effects in industries with low labor turnover such as manufacturing than in fluid sectors such as restaurants. Overall, this suggests that workers who lose their jobs in declining low-turnover sectors will suffer more adverse consequences, in particular if much of their human capital is industry-specific. 

In order to quantify the conditions in each industry, we first construct time-consistent industry indicators at the 3-digit level. This is a non-trivial endeavour as the codes changed three times during our sample period. We start from the SNI2002 codes that our raw data provide for the period 2002 to 2010. Next, we rely on the SNI2002 code reported in 2010 and use it for all later years for those establishments that continue to exist beyond 2010, and conversely use the data from 2002 for establishments that existed prior to that. We refer to these overlapping establishments as stayers. We then use the modal overlap between SNI2002 (as imputed for stayers) and the current codes (SNI69, SNI92 and SNI2007) to fill in SNI2002 codes for non-stayers. This works particularly well for the years and sample we study, as almost all closing establishments existed in 2010 (due to the 3-year tenure requirement) and since most codes remained unchanged between SNI92 and SNI2002.    

We measure \textit{Industry wage} premia in each 3-digit industry as average earnings in the industry, residual on years of schooling, 2-digit education field, gender, immigrant background and a full set of age dummies. Furthermore, we also include average \textit{Churn} and \textit{Reallocation} rates for each industry. We follow conventions from \cite{Burgess2000}, and define establishment-level churning as the number of workers who are hired or separated beyond what was needed for the actual change in employment between two adjacent years. The churning rate is measured as churning relative to the average employment during the two years.\footnote{Thus, churning is [(Hires+Separations)-abs(Emp(t)-Emp(t-1))]/[Emp(t)/2+Emp(t-1)/2] at the establishment-year level.} Similarly, we calculate excess reallocation in each industry as the excess creation and destruction of jobs over what was needed to adjust industry employment. The reallocation rate is measured relative to the average employment in the two years.\footnote{Thus, reallocation is [(Job Creation+Job Destruction)-abs(Emp(t)-Emp(t-1))]/[Emp(t)/2+Emp(t-1)/2] at the industry-year level.} We then aggregate the establishment-year churning rates and the industry-year reallocation rates to industry-level averages which are constant over time. When computing these numbers, we take care to exclude workers that satisfy the conditions for false closures as discussed above. Note also that the scale of the measures (relative to average employment in the two years) is in approximate percentages, but with a maximum of 2 and a minimum of -2.  

We also follow the conventions from this literature when computing \textit{Employment growth over the past 10 years} in each industry. Thus, for displacement events in the year $t=-1$, we define $Trend_{Ind,t}=(Emp_{Ind, t-1}-Emp_{Ind, t-10})/(Emp_{Ind, t-1}/2+Emp_{Ind, t-10}/2)$ which takes the value 2 for newly emerging industries and the value -2 for disappearing industries. This metric bounds cases where some very small industries experience extreme changes (including exits and entries) during the sample period. Similarly, we calculate the \textit{Industry-specific business cycle} as the change in employment between year $t=-1$ and $t=0$ using the same metric.

We also add dummy variables for closures in the \textit{Manufacturing industry}, as this industry has been a focus of the job displacement literature since \cite{JLS1993}. Furthermore, we add a dummy for closures within \textit{Education, health and public administration}, as the labor markets in these (mostly public sector) industries tend to experience a constant shortage of workers.

\subsection{Location Characteristics} 
As with industries, there is ample evidence suggesting that local labor market conditions have causal effects on workers' outcomes. This is also the case in Sweden, where wages seem to react to local labor market shocks \citep{Hakkinen2019}.

Our local variables are measured at the level of local labor markets, which are an aggregation of municipalities. These are constructed by Statistics Sweden based on commuting patterns. 

For each local labor market, we measure the \textit{Local unemployment} rate as the number of residents who are registered with the public employment service (a prerequisite for receiving benefits from either the unemployment insurance system or from the municipal welfare system) divided by the size of the local labor force (sum of the registered unemployed and the number of employed as described above). \textit{Population density} is also measured at the local labor market level. The \textit{Concentration of local employment across 3-digit industries} is measured using an HHI index.

In addition, we measure local exposure to the industry characteristics discussed above. These shift-share/Bartík-style variables are calculated by multiplying each industry's employment share in the local labor market (by year) with the characteristics of that industry and then summing over the industries within the local labor market. In this way, we define the \textit{Manufacturing employment share}, \textit{Average long-term industry trend}, \textit{Average industry business cycle}, \textit{Average industry churn} and \textit{Average industry reallocation}. 

It is important to note the fundamental difference between these variables and their industry-level counterparts. Whereas the industry-level variables measure characteristics in the industry from which the worker was displaced (and is potentially tied to, if changing industry is costly), the local labor market counterpart measure how exposed the worker is to these attributes if searching at random at the local labor market (which should be more pertinent for workers who are restricted in their mobility).

\subsection{Aggregate Conditions}
To capture aggregate conditions in the economy, which may both impact the average size of earnings losses, as well as have heterogeneous effects on different worker groups, we include the calendar \textit{Year} ($t=0$) and the \textit{National unemployment rate} in the year $t=1$. The national unemployment rate captures the aggregate business cycle conditions displaced workers experience after job loss. These have been shown to be an important correlate of earnings losses by \cite{DavisWachter}.
\clearpage

\newpage \clearpage

\section{Online Appendix: Additional Results}


 \begin{center}
       
 \footnotesize
    
\begin{longtable}{p{8cm}*3c} 
\caption{Sample statistics for displaced and non-displaced workers}\label{t:desc}
    \endfirsthead
    \endhead
      \toprule 
   &Controls&Controls&Displaced\\
   &(unmatched)&(matched)&(unmatched)\\
    & (1) & (2) & (3)\\
  \midrule
N workers   &28,914,261& 556,824& 185,627\\
N establishments& 231,958& 120,597&  22,005\\
&&&\\
\textbf{Covariates} &&&\\
{\emph{Demographics}} &&&\\
Age         &      44.113&      42.146&      42.157\\
Female      &       0.478&       0.346&       0.347\\
Born abroad &       0.097&       0.129&       0.131\\
Two foreign born parents&       0.028&       0.032&       0.032\\
Native, not birth county&       0.274&       0.249&       0.249\\
&&&\\
{\emph{Family}} &&&\\
Married     &       0.645&       0.589&       0.588\\
Divorced    &       0.101&       0.099&       0.099\\
Children (\# living at home)&       0.780&       0.749&       0.746\\
School-aged children&       0.524&       0.470&       0.467\\
Share household inc.&       0.714&       0.754&       0.754\\
Past mobility (\# location moves in last 10 years)&       0.164&       0.212&       0.211\\
&&&\\
\emph{Human capital} &&&\\
Schooling (years of education)&      12.160&      11.466&      11.470\\
Exp. (years employed in last 10)&       9.229&       8.913&       8.908\\
Tenure      &       6.624&       5.896&       5.890\\
Industry tenure&       7.927&       7.137&       7.138\\
Earnings(rank)&       0.500&       0.497&       0.497\\
Earnings in t-2&       0.994&       1.013&       1.014\\
Earnings in t-3&       0.951&       0.961&       0.962\\
STEM education&       0.097&       0.108&       0.109\\
Education in licensed field&       0.238&       0.070&       0.071\\
Education specificity&       0.565&       0.470&       0.470\\
&&&\\
\emph{Lost job characteristics} &&&\\

Establishment size&     520.493&     177.714&     185.309\\
Establishment size trend&       0.025&      -0.020&      -0.022\\
Establishment wage premium&       0.001&      -0.032&      -0.033\\
Job routineness&       0.515&       0.557&       0.557\\
Industry matched to education&       0.560&       0.377&       0.376\\
Size of displacement&       0.170&       0.157&       0.158\\
Manager     &       0.033&       0.026&       0.027\\

&&&\\
\emph{Industry characteristics} &&&\\

Employment trend&       0.052&       0.079&       0.081\\
Employment cycle&       0.005&      -0.001&      -0.001\\
Industry wage premium&       0.006&       0.028&       0.027\\
Churning    &       0.223&       0.209&       0.209\\
Reallocation rate&       0.131&       0.140&       0.140\\
Manufacturing&       0.230&       0.363&       0.366\\
Education, health, admin&       0.383&       0.078&       0.079\\

&&&\\
\emph{Location characteristics} &&&\\

Population density&      82.472&      86.275&      86.295\\
Unemployment&       0.086&       0.086&       0.086\\
Concentration (HHI industries)&       0.031&       0.031&       0.031\\
Manufacturing (share of local emp.).&       0.189&       0.186&       0.186\\
Average industry trends&       0.081&       0.079&       0.079\\
Average industry cycle&       0.007&       0.007&       0.007\\
Average industry churning&       0.224&       0.226&       0.226\\
Average industry reallocation&       0.147&       0.148&       0.148\\

&&&\\
Year        &   2,005.881&   2,005.580&   2,005.580\\
National unemployment&       7.451&       7.432&       7.432\\
\bottomrule
\multicolumn{4}{{m{15cm}}}{\scriptsize \textit{Note:} Statistics for the non-displaced (controls) before matching (Column 1) and after matching (column 2), and for the displaced workers (Column 3). Mean values of all covariates used in the analyses. Details for all covariates are in Appendix A. 
}
\end{longtable}
   \end{center}


\begin{table}[H]
\caption{RATE estimates of GRF model and different simple targeting rules on the test set}\label{Tab4_RATE}
\footnotesize
\begin{centering}
\begin{tabular}{lcc}
\toprule
           &Qini coefficient (x100)&    SE(x100)\\
\midrule
GRF         &       -3.56&          .2\\
Age         &       -1.74&          .2\\
Schooling   &       -1.47&          .2\\
Manufacturing&       -1.52&          .3\\
Population density&       -1.08&          .3\\
Manufacturing \& age&       -2.43&          .3\\
Manufacturing \& schooling&       -2.16&          .3\\
\addlinespace
\bottomrule
\addlinespace
  \multicolumn{3}{p{0.98\textwidth}}{\scriptsize \textit{Note}: \textit{GRF}: workers with the lowest estimated CATEs; \textit{Age}: oldest workers; \textit{Schooling}: least educated workers; \textit{Manufacturing}: manufacturing workers at random and then other workers; \textit{Population density}: least dense locations; \textit{Manufacturing and age}: oldest manufacturing workers; \textit{Manufacturing and schooling}: least educated manufacturing workers.}
\end{tabular}
\end{centering}
\end{table}

\newpage \clearpage

  
\begin{figure}[H]
	\caption{Distribution of relative earnings five and ten years after displacement}
	\label{Rel_alt}
         \begin{center}
      \includegraphics[width=\textwidth]{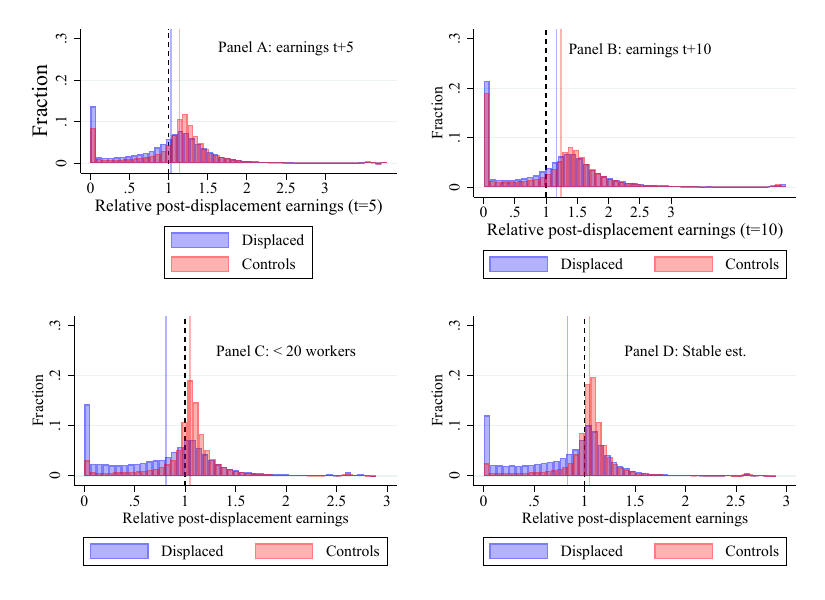}
      \end{center}
   \footnotesize \textit{Note}: Panels A and B show the distributions of relative earnings between one year before displacement and five years after displacement (Panel A) and ten years after displacement (Panel B). Panels C and D show distributions of relative earnings in $t=1$ for displaced and matched controls in sub-samples with less than 20 workers (Panel C) and with less than 10 percent change in employment between t-3 and t-1 (Panel D), similar to Figure \ref{fig1_ATE}B in the paper. Solid lines indicate group means, the dashed lines indicate unchanged nominal earnings. 
  \end{figure}

\newpage \clearpage


  \begin{figure}[H]
	\caption{Robustness of ATE estimates for the top and bottom CATE deciles for subgroups, alternative control groups, and alternative estimation methods}
	\label{App:FigRobMatch}
         \begin{center}
      \includegraphics[scale=1.1]{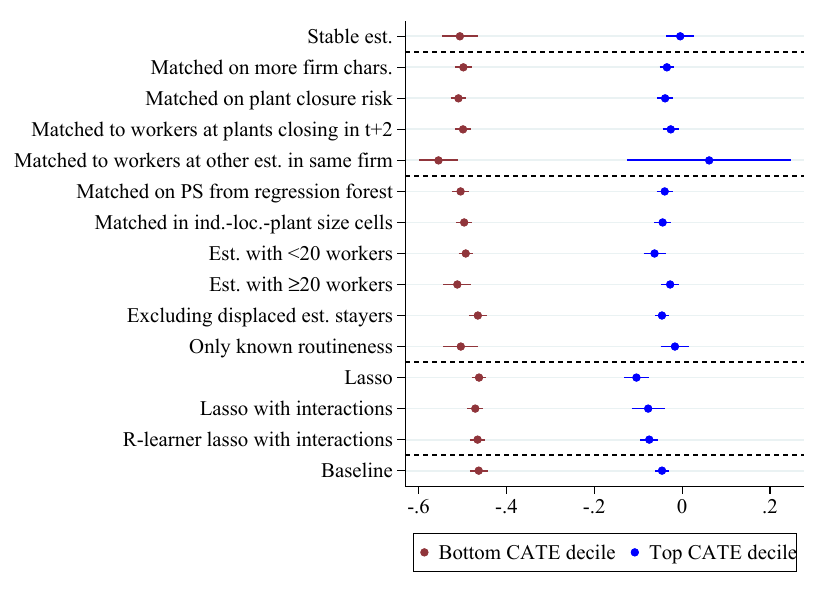}
      \end{center}
   \footnotesize \textit{Note}: ATE estimates for individuals in the top and bottom deciles of CATEs for different subgroups, using different matching approaches or different estimation methods. In the case of subgroups, displaced from that subgroup are re-matched to controls to ensure that all matched controls satisfy the subgroup restrictions. \textit{Stable est.}: establishments with absolute employment growth $<$ 10\% from $t=-3$ to $t=-1$. \textit{Matched on more firm chars.}: in addition to all variables used for matching in the baseline, matched on the wage quantile, hire share and leaver share in the firm the establishment belongs to (measured as the firm's average in the years $t=-5$ through $t=-1$). \textit{Matched on plant-closure risk}: matching based on predicted plant-closure risk, using all industry and location variables, establishment size, trend in establishment size, plant wage level, plant share of location-industry, firm wage quantile, hire share and leaver share to predict the risk. \textit{Matched to workers at plants closing in t+2}: control group consists of workers employed in $t=-1$ at plants which close in $t=2$. \textit{Matched on PS from regression forest}: propensity scores for matching estimated using regression forest instead of logit. \textit{Matched in ind.-loc.-plant size cells}: matching within cells defined by combinations of metropolitan/non-metropolitan local labor markets, manufacturing/education, health, public administration/other non-manufacturing industries, establishments with 5-20/21-100/100+ employees. \textit{Est. with $<20$ workers}: establishments with $<20$ workers in $t=-1$. \textit{Est. with $\geq20$ workers}: establishments with $\geq20$ workers in $t=-1$. \textit{Excluding displaced est. stayers}: excluding individuals who are still employed at a closing establishment in $t=1$. \textit{Only known routineness}: only individuals whose occupations are known and hence routineness is non-imputed. 
   \textit{Lasso}: lasso using displacement interacted with the variables. 
   \textit{Lasso with interactions}: lasso using displacement interacted with the variables, their squares and within-block interactions. \textit{R-learner lasso with interactions}: R-learner lasso based on the variables, their squares and within-block interactions.

 \end{figure}

\clearpage

  \begin{figure}[H]
	\caption{ATE:s on earnings in t+3 and on cumulative earnings in t+1 to t+7}
	\label{App:Fig_grfs_t3_t5_t7}
         \begin{center}
      \includegraphics[width=\textwidth]{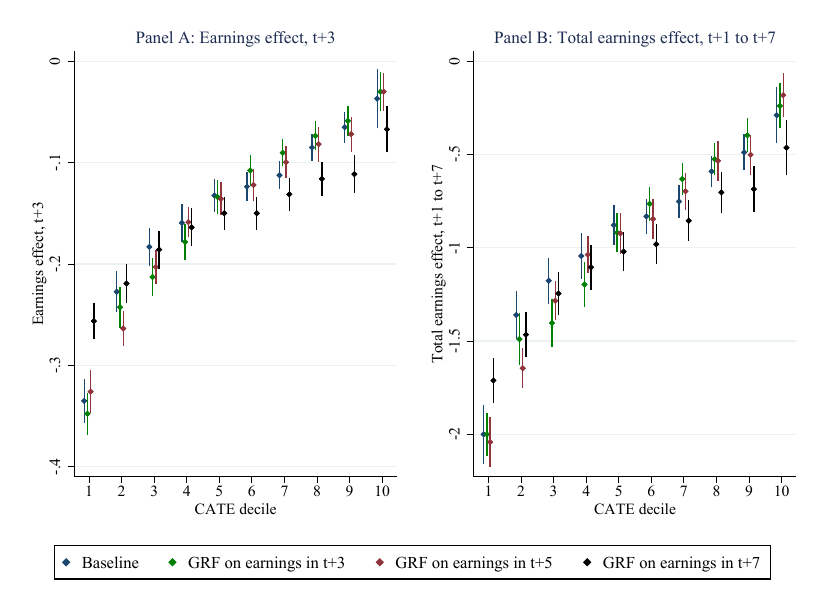}
      \end{center}
   \footnotesize \textit{Note}: \textit{Panel A}: ATE:s (differences between displaced and control workers) in terms of earnings three years after displacement. \textit{Panel B}: ATE:s (differences between displaced and control workers) in terms of cumulative earnings from one to seven years after displacement. Workers sorted into deciles of CATE:s based on GRF models estimated on earnings in $t+1$ (baseline), $t+3$, $t+5$ and $t+7$ respectively. 95 percent confidence intervals.
 \end{figure}

 \begin{figure}[H]
	\caption{Effects of displacement on employment over time}
	\label{f:emp_potential}
         \begin{center}
      \includegraphics[width=\textwidth]{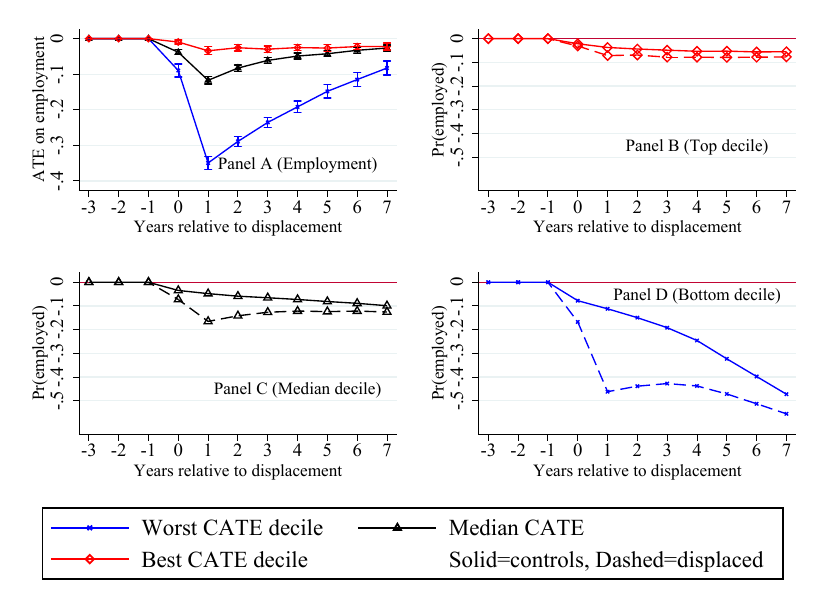}
      \end{center}
   \footnotesize \textit{Note}: The figure replicates figure \ref{fig3_ATE_andY1Y0} in the paper, but with employment as the outcome instead of earnings. It shows statistics for three decile groups of the CATE distribution. The \say{median} group straddles the median (i.e. it contains the $10th$ and $11th$ ventile). Panel A shows the ATE over time, similar to figure \ref{fig1_ATE}, but separately for the decile groups. Point estimates and 95 percent confidence interval with standard errors clustered at the establishment in $t=-1$. Panels B--D show the underlying employment trajectories for displaced and matched controls within each decile group for the top, bottom and median deciles respectively. Only workers observed in each of the periods $t=-3$ to $t=7$ are included (this entails excluding the years 2011-2014). Sample restrictions ensure that employment is equal to $1$ for all until $t=-1$. 
  \end{figure}


  \begin{figure}[H]
	\caption{Differences in characteristics across CATE quartiles and deciles}
	\label{fig_interdecile}
         \begin{centering}
      \includegraphics[width=\textwidth]{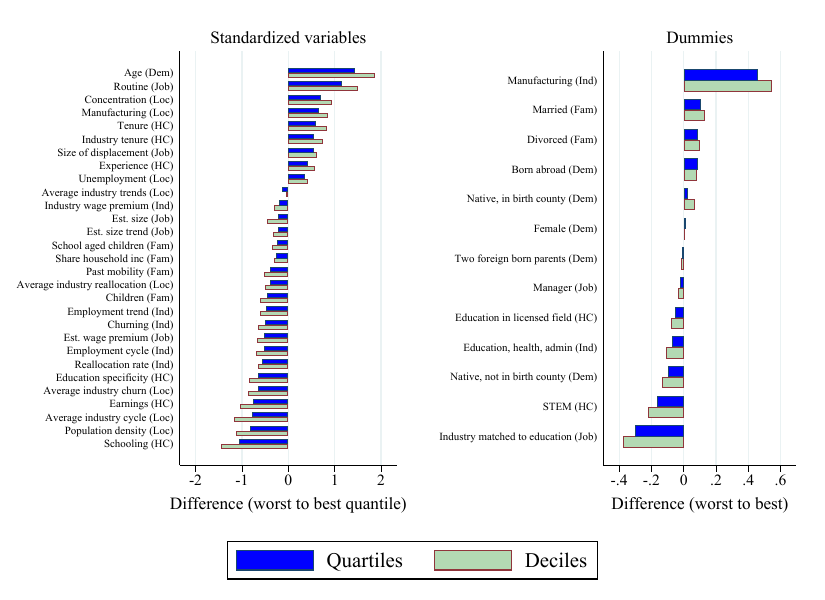}
      \end{centering}
   \footnotesize \textit{Note}: The figure shows differences in characteristics between individuals in the lowest and highest quartiles and deciles of CATE:s, using the training data set. CATE:s estimated using 5-fold estimation and ranking done within each fold. The left-hand panel contains standardized (mean $0$ and standard deviation $1$) continuous variables and the right-hand panel contains dummy variables. Blue bars indicate quartiles and green bars indicate deciles. 
  \end{figure}

\clearpage

  \begin{figure}[H]
	\caption{CATE across combinations of age and schooling}
	\label{Fig:AppCountourCATE}
         \begin{center}
      \includegraphics[width=\textwidth]{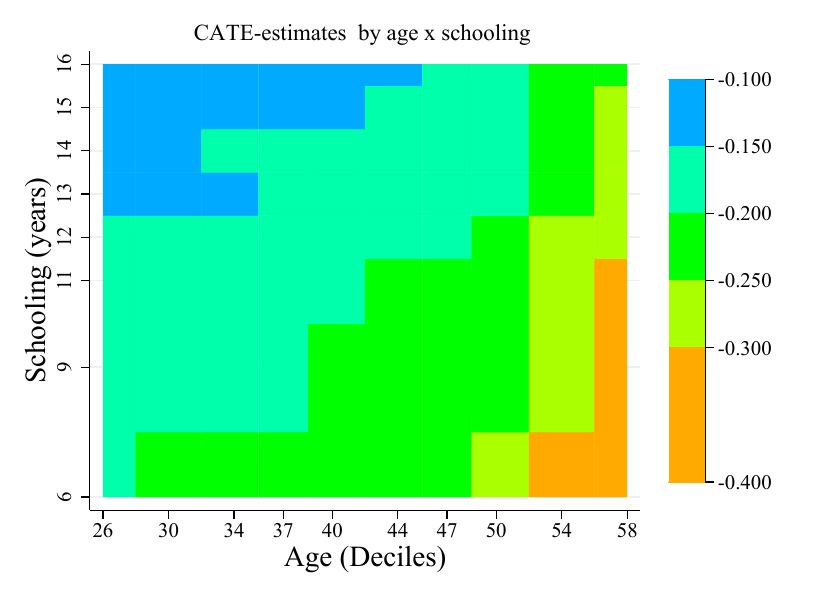}
      \end{center}
   \footnotesize \textit{Note}: The figure replicates Figure \ref{fig6_contour}A (and B) in the paper, but using CATE instead of ATE (or AIPW). It divides training set workers into cells by age and schooling. Schooling has been aggregated to 8 groups by pooling the few with 10 years of schooling together with those with 9 years of schooling, and by letting the top group include all with 16 or more years of schooling. Age is defined in deciles among the displaced and the x-axis shows the median in each age group. Colors indicate the size of point estimates. It shows estimated CATE:s by combinations of age and schooling. 
 \end{figure}

\clearpage

  \begin{figure}[H]
	\caption{Heterogeneity within extreme combinations of age and schooling}
	\label{fig7_Extreme}
         \begin{center}
      \includegraphics[width=\textwidth]{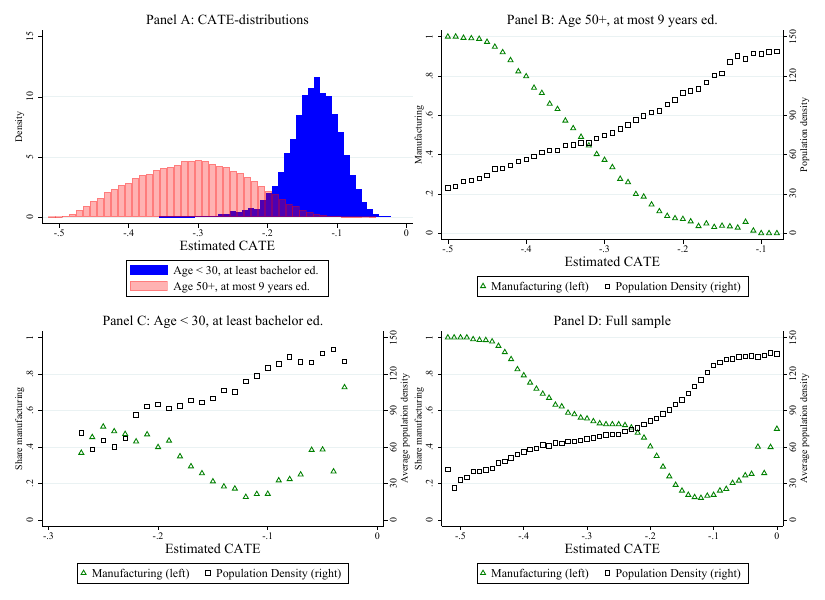}
      \end{center}
   \footnotesize \textit{Note}: Training set workers who are (a) younger than 30 and have at least 15 years of education or (b) older than 50 and have at most 10 years of education. Panel A: Histograms of 5-folds GRF CATEs within these groups. Panels B-C: Share of manufacturing workers and average population density within each CATE cell for the two groups.  Panel D repeats this for the full sample. Cells defined as CATE bins of one percentage point. Bins containing fewer than ten workers are dropped.  
 \end{figure}
\clearpage


\begin{figure}[H] 
\caption{Robustness analyses: Displacement effects across and within establishments using AIPW}
\label{Appfig_within_aipw}
\begin{centering}
    \includegraphics[width=\textwidth]{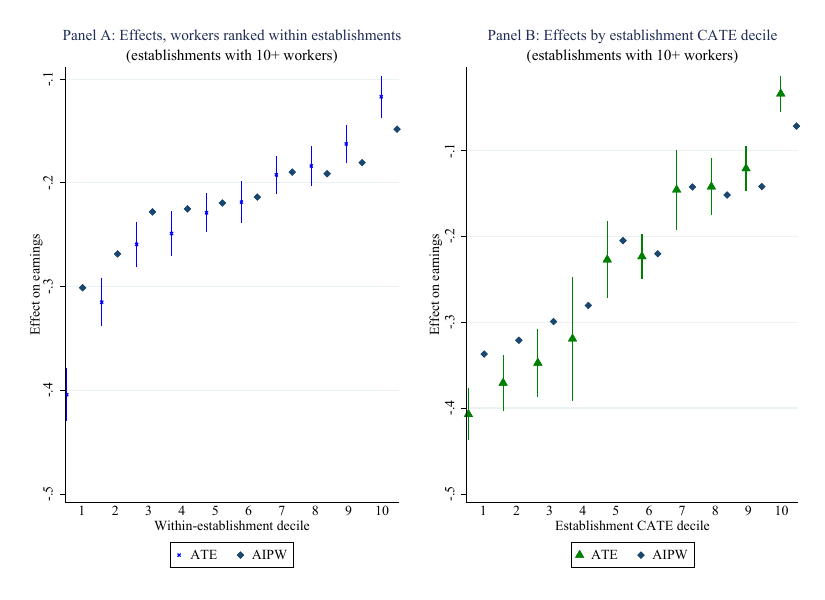}
\end{centering}
   \scriptsize \textit{Note}: The figure shows robustness analyses using both ATE:s and average AIPW scores in the respective CATE deciles. Panel A shows ATE/AIPW estimates when displaced workers are ranked based on their within-establishment CATE. Panel B shows ATE/AIPW estimates when displaced workers are ranked based on the CATE of their co-workers (defined as the leave-out mean for the workers at the establishment and then averaging over the individuals in the CATE decile.) This analysis excludes establishments with fewer than 10 displaced workers. Controls are allocated to the same decile sample as the treated workers they were matched to.
\end{figure}

\clearpage


\begin{figure}[H] 
\caption{Robustness analyses: Displacement effects across and within establishments for additional outcomes}
\label{Appfig_within_out}
\begin{centering}
    \includegraphics[scale=1.3]{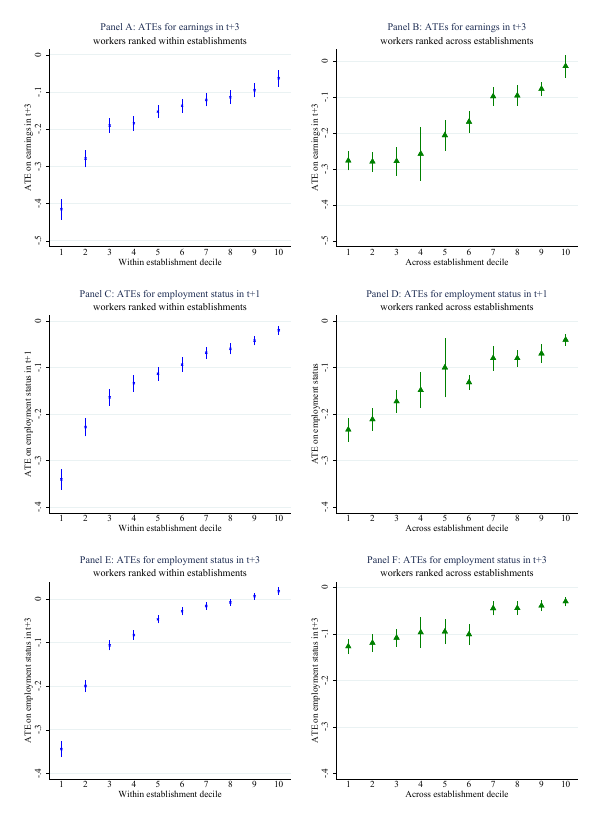}
\end{centering}

   \scriptsize \textit{Note}: The figure shows robustness analyses using employment status in $t+1$, earnings in $t+3$ and employment status in $t+3$ as outcomes. Panels A, C and E show ATE estimates when displaced workers are ranked based on their within-establishment CATE. Panels B, D and F show ATE estimates when displaced workers are ranked based on the CATE of their co-workers (defined as the leave-out mean for the workers at the establishment and then averaging over the individuals in the CATE decile.) Panels A and B exclude establishments with fewer than 10 displaced workers. Controls are allocated to the same decile sample as the treated workers they were matched to.
\end{figure}

\clearpage

\begin{figure}[H] 
\caption{Robustness analyses: Displacement effects across and within establishments by AKM plant estimates}
\label{Appfig_within_AKM}
\begin{centering}
    \includegraphics[width=\textwidth]{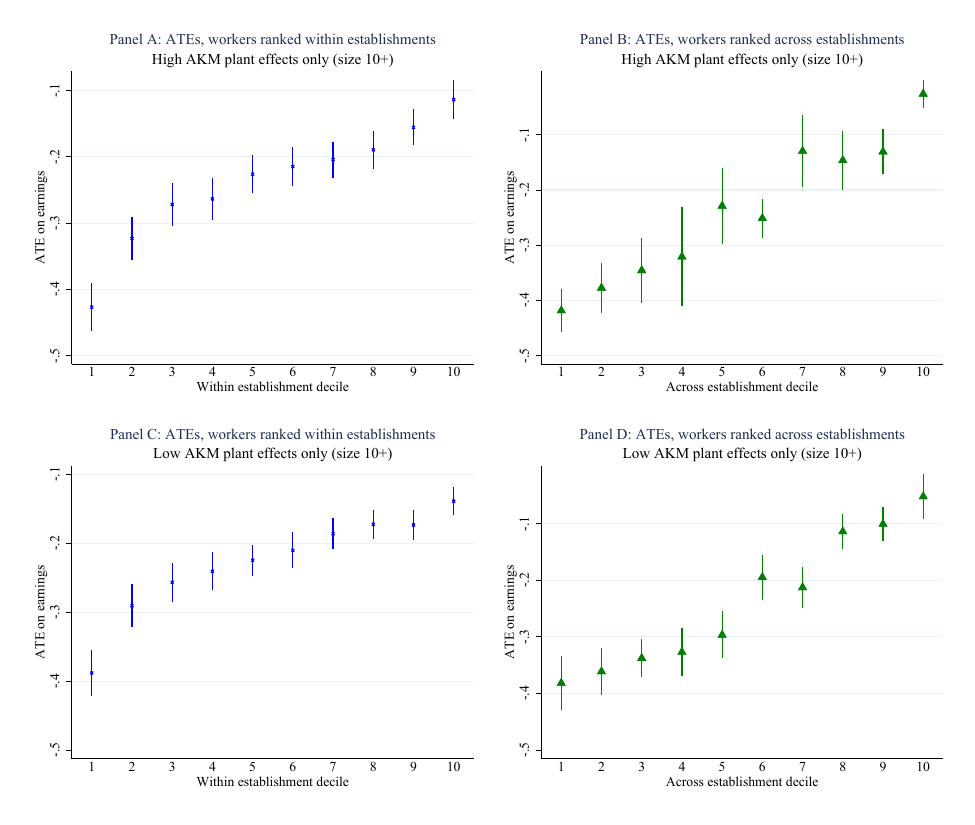}
\end{centering}
   \scriptsize \textit{Note}: Panel A (high AKM-plant effects) and C (low AKM-plant effects) shows ATE estimates when displaced workers are ranked based on their within-establishment CATE. Panel B  (high AKM-plant effects) and D (low AKM-plant effects) shows ATE estimates when displaced workers are ranked based on the CATE of their co-workers (defined as the leave-out mean for the workers at the establishment and then averaging over the individuals in the CATE decile.) All panels exclude establishments with fewer than 10 displaced workers. Controls are allocated to the same decile sample as the treated workers they were matched to. AKM-estimates \citep{AKM1999} are derived from estimates on rolling 5-year samples ending in the year prior to the event ($t=-1$). Models have fixed establishment ("plants") effects, fixed person effects, year dummies, and third order polynomial in age (in deviation from 45), separately for three education groups (less than high school, high school, more than high school). The outcome is log earnings conditional on employment. Sample is split into high vs. low plant-effects samples by the median plant effect within each displacement cohort.  
\end{figure}

\clearpage


\begin{figure}[H] 
\caption{CATE across and within establishments}
\label{Appfig_within}
\begin{centering}
    \includegraphics[width=\textwidth]{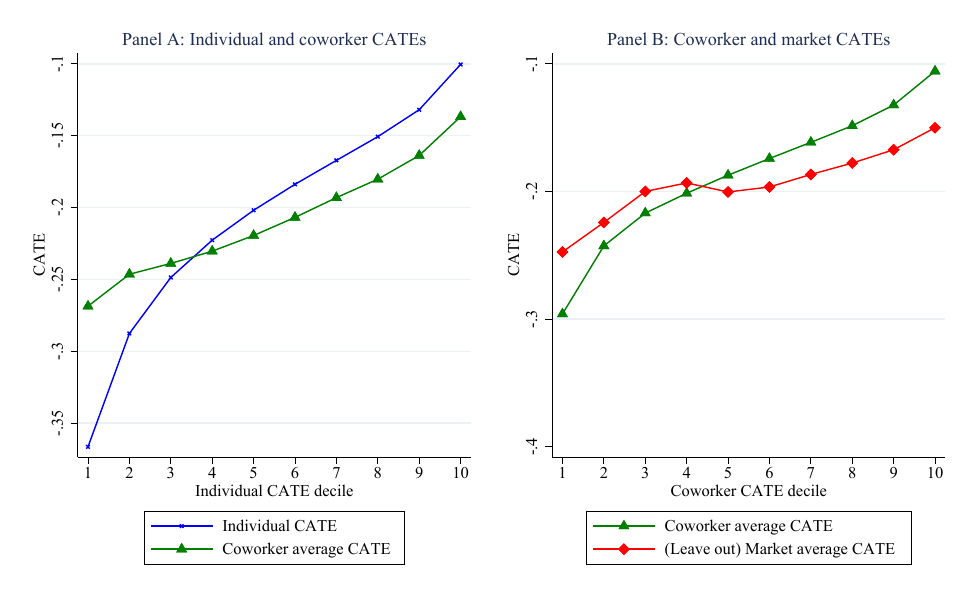}
\end{centering}
   \scriptsize \textit{Note}: Panel A shows statistics for displaced workers in the main (training) set by CATE decile (5-fold estimation and ranking done within each fold). It reports average individual CATE and the average CATE for the coworkers of the individuals in each CATE decile taken as a leave-out mean across the workers at the establishment and then averaging over the individuals in the CATE decile. Panel B shows average coworker CATE and the average CATE for other closures in the same industry, location and year by coworker CATE decile. Establishments with less than 10 displaced workers are discarded in Panel A. Workers from markets with only one event are discarded in Panel B.
\end{figure}

\clearpage


\begin{figure}[H] \caption{Heterogeneity within and across establishments, interactions}
\label{fig_ATEContour}
    \begin{center}
    \includegraphics[scale=0.6]{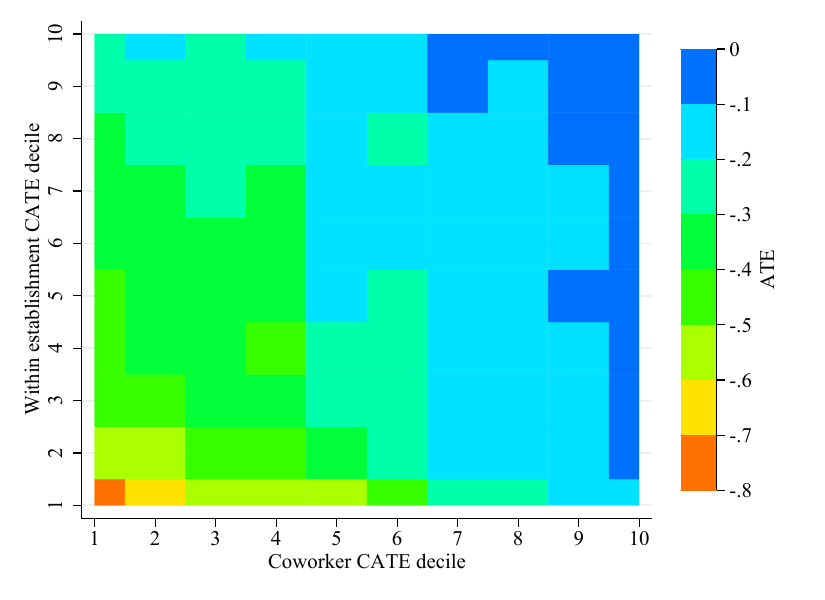}
   \end{center}
   \footnotesize \textit{Note}: The figure divides training set workers by combinations of coworker 5-fold CATE decile and within-establishment decile. Coworker CATE decile is determined by the ranking of the leave-out mean CATE at the establishment. For the within-establishment CATE, individuals are ranked by their CATE:s within the establishment. The figure restricts to establishments with at least 10 workers. Colors indicate the size of point estimates of the CATE within each cell.    
\end{figure}


  \begin{figure}[tbh]
	\caption{Insurance across the CATE distribution}
	\label{fig4_dispinc}
         \begin{centering}
      \includegraphics[width=\textwidth]{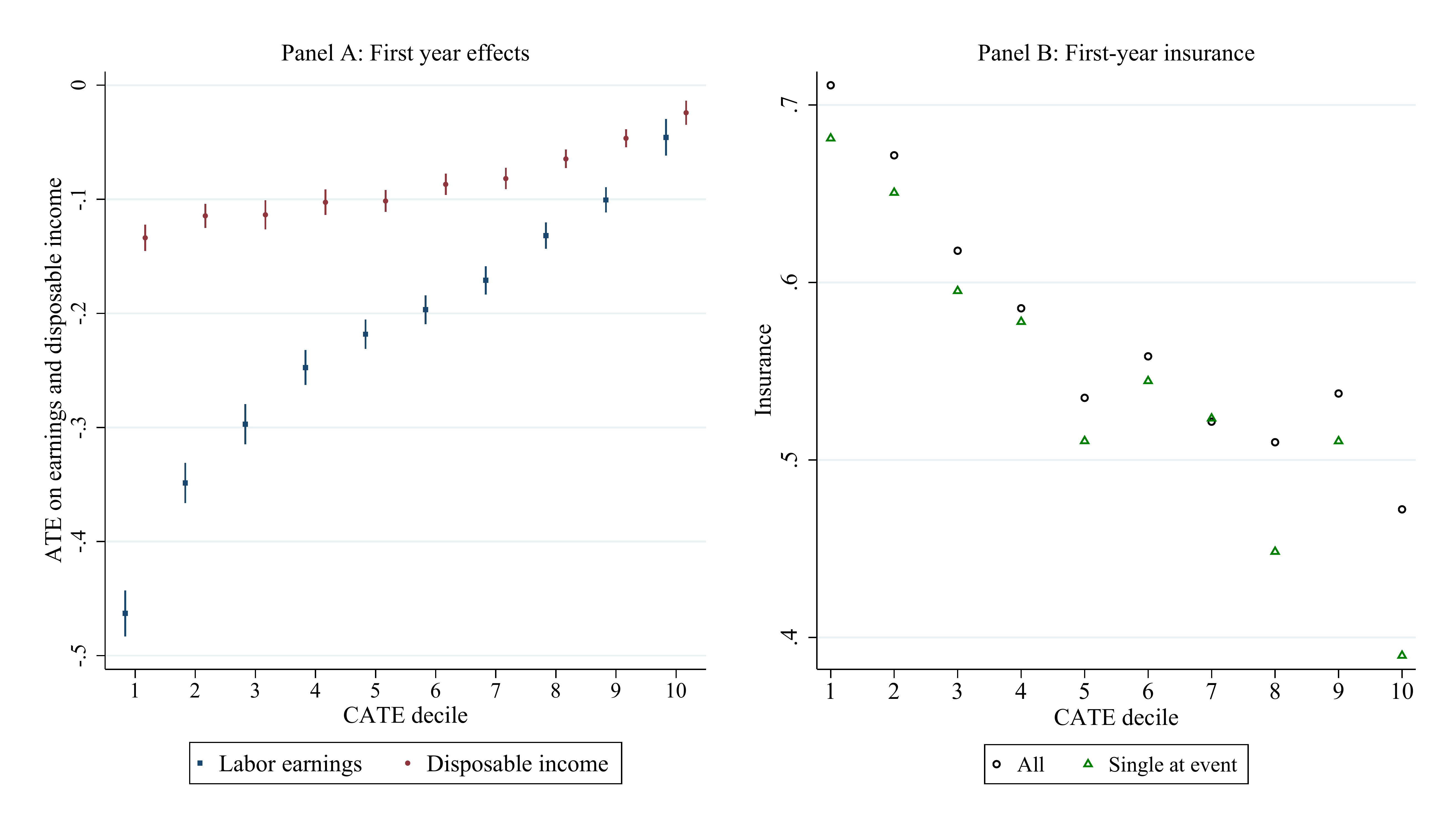}
      \end{centering}
   \footnotesize \textit{Note}: The figure shows outcomes when dividing the training data set into deciles based on CATE:s estimated with 5-fold estimation. Panel A shows ATE estimates on labor earnings (as in figure \ref{fig2_CATE_ATE}) and corresponding ATE estimates for individualized disposable income. Point estimates and 95 percent confidence interval with standard errors clustered at the establishment level. Panel B shows the implied degree of insurance, defined as the difference between the ATE on disposable income the ATE on labor earnings,  divided by the ATE on labor earnings. It provides separate estimates for the full population of displaced and for singles without children (in the year before displacement).  

  \end{figure}

\clearpage


  \begin{figure}[tbh]
	\caption{Differences in characteristics across CATE quartiles for disposable income}
	\label{fig4_dispinc2}
         \begin{centering}
      \includegraphics[width=\textwidth]{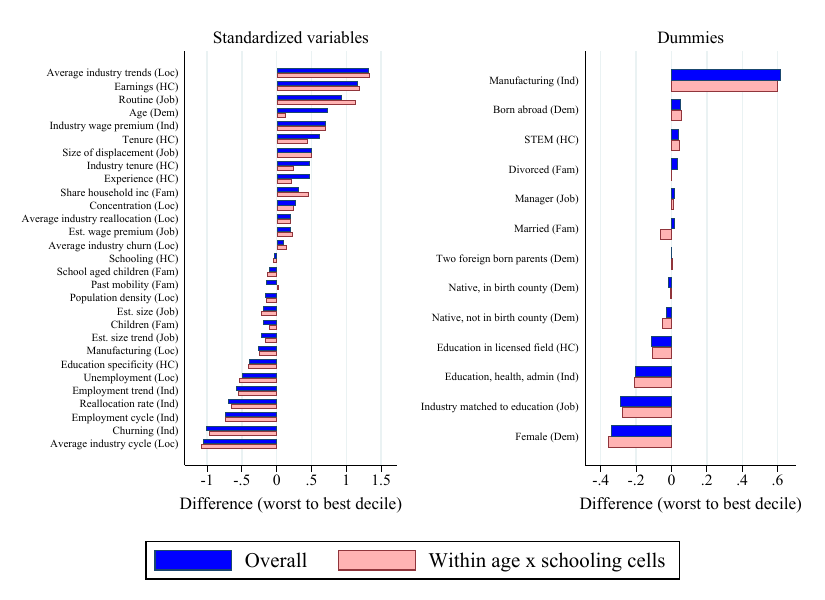}
      \end{centering}
   \footnotesize \textit{Note}: The figure divides the sample by the CATE estimates for disposable income. It shows differences in characteristics between individuals in the lowest and highest quartiles and deciles of CATE:s, using the training data set. CATE:s estimated using 5-fold estimation and ranking done within each fold. The left-hand panel contains standardized (mean $0$ and standard deviation $1$) continuous variables and the right-hand panel contains dummy variables. Blue bars are for the overall quartiles and red bars cover the highest and lowest quartiles of CATE within each combination of 8 schooling and 10 age categories
  \end{figure}

\clearpage



\section{Online Appendix: Partial Effects of Market Conditions}

This appendix describes the most important industry and location variables in more detail. Isolating the importance of market-level factors is not trivial since workers are likely to be sorted across regions and industries. For that reason we use the GRF to estimate partial dependence functions, holding the individual characteristics at their empirical levels and sequentially rotating across all observed sets of market characteristics within the same year. This way, we capture the predicted role of aggregate conditions across the full distribution of displaced workers, making full use of the GRF's non-linear nature. Because market-level factors are correlated, we split locations and industries into groups, and then characterize effects and attributes for combinations of these groups.

Formally, we divide the vector of characteristics $X$ into parts related to the location ($X^L$), the industry ($X^S$), or the displaced worker and lost job ($X^I$). Worker $i$ was displaced from location $l(i)$ and industry $s(i)$, with $X_{l(i)}^L$ and $X_{s(i)}^S$ being the corresponding location and industry characteristics. Then we compute: 
\begin{equation}\label{eq:ReLocations}
    \tau_l^L=\frac{1}{N} \sum_{i=1}^N\text{CATE}(X^L=X_l^L,X_{s(i)}^{S},X_i^I), \quad\tau_s^S=\frac{1}{N} \sum_{i=1}^N\text{CATE}(X_{l(i)}^L,X^S=X_s^S,X_i^I),
\end{equation}
where $N$ is the number workers in the sample. $\tau_l^L$ and $\tau_s^S$ are the average CATEs if all N workers experienced the conditions in location $l$/industry $s$. We use these predictions to rank industries and locations by $\tau^L$ and $\tau^S$ respectively. 

We classify workers using a similar strategy. Here, we hold worker characteristics fixed and combine them with each observed combination of location and industry conditions. Then for each worker $i$ we compute:
\begin{equation}\label{eq:WorkerRes}
    \tau_i^I = \frac{1}{N^L \times N^S} \sum_{l=1}^{N^L}\sum_{s=1}^{N^S} \text{CATE}(X_l^L,X_{s}^{S},X_i^{I}), 
\end{equation}
where $N^L$ and $N^S$ are the number of locations and industries respectively. $\tau_i^I$ only reflects worker-level characteristics since we average over the same set of location and industry conditions for all workers (by year). 

In the end, we compute how ATEs, and characteristics, vary with (combinations of) quartiles of $\tau^L$, $\tau^S$ and $\tau^I$. This strategy allows us to reduce the dimensionality of the heterogeneity and study how different types of heterogeneity interact. 

\subsection{ATE Estimates by Partial Market and Worker Quartiles}
Table \ref{t: shuffle} shows estimated ATEs for different cuts of the data, using combinations of $\tau_l^L$, $\tau_s^S$ and $\tau_i^I$. Panel A shows estimates for the lowest (Column 1) and highest (Column 2) quartile of worker-level effects ($\tau_i^I$). Column 3 shows the inter-quartile differences, with standard errors in Column 4. The top row shows that workers in the lowest quartile of $\tau_i^I$ experience $26$ pp larger losses than the top quartile. 

In the rows below, we zoom in on workers exposed to different combinations of location and industry conditions. We see significant differences in ATEs across worker types within each set of market conditions (ranging from $0.30$ to $0.16$). Market conditions are particularly important predictors for workers with poor individual characteristics.\footnote{See Appendix Figure \ref{fig_ATEContour} for similar results on the importance of establishment heterogeneity for low ranked workers.} The ATEs vary nearly as much when comparing across market (industry and location) conditions for the lowest worker quartile ($25$ pp difference, comparing across rows) as when comparing across worker quartiles within the worst market (industry and location) quartile setting ($30$ pp).  

In Panels B and C, the columns instead represent location and industry (defined at the 3-digit level) quartiles, respectively. Here, the different rows represent worker types.\footnote{Standard errors for the cross-quartile differences are clustered at location (Panel B) and industry (Panel C) since the number of locations and industries is small relative to the sample.} Panel B shows that the location matters for the size of the displacement effect even when holding the worker quartile ($\tau_i^I$) fixed (interquartile differences are $6$ to $12$ pp depending on type). Workers also suffer significantly larger effects if displaced from \say{bad} industries (Panel C), in particular if they belong to the bottom worker quartile. In this group, estimated losses are $19$ pp larger if displaced in a low-quartile industry as compared to a high-quartile industry. 

Market conditions should matter less (or not at all) if workers are mobile. We therefore estimate how displacement events affect mobility across locations and industries. Panel D shows that workers are more likely to change location if displaced under worse conditions. The estimated effects on location mobility are small, partly because location mobility overall is very low, but the results show that this is true even for workers displaced in very bad locations. The results also indicate that job loss leads to more mobility across locations among workers who are more resilient. 

Panel E shows that workers are more likely to move away from their (1-digit) industry if they are displaced from an industry with large displacement effects. This analysis is performed on the endogenous subsample of workers who find new employment, which warrants some caution. With this caveat in mind, the results suggest that cross-industry mobility due to job loss is more common among workers in industries where effects are larger; 33 (23) percent move because of displacement in industries in the worst (best) industry quartile. The magnitudes here are much more substantial than for location mobility thorughout, which is natural since the old job disappears, but (other) ties to the location remain. Even though we find that industry mobility responds to displacement during distressful conditions, the fact that industry characteristics in general correlate so strongly with displacement effects suggests that the degree of industry mobility is insufficient to offset the negative effects of being displaced in a bad industry.

\begin{table}[H]
\footnotesize
\caption{Earnings and mobility effects, by partial market and worker characteristics}\label{t: shuffle}
\begin{centering}
\begin{tabularx}{\textwidth}{l zz zz}
\toprule
& (1) & (2) & (3) & (4) \\
&Worst&Best&Interquartile&Standard Error\\
&Quartile&Quartile&Difference&of Difference\\
\addlinespace
\midrule
\multicolumn{5}{l}{\textbf{Earnings Estimates}}\\
\midrule
\multicolumn{5}{l}{\textit{Panel A: Worker Quartiles, earnings effects}}\\
All Market types & -0.374 & -0.100 & -0.274 & 0.008 \\
\addlinespace
Worst Market quartiles & -0.485 & -0.141 & -0.343 & 0.076 \\
Median Markets 		& -0.360 & -0.120 & -0.240 & 0.027 \\
Best Market quartiles & -0.234 & -0.068 & -0.166 & 0.021 \\
\midrule
\multicolumn{5}{l}{\textit{Panel B: Location Quartiles, earnings effects}}\\
All worker types & -0.296 & -0.171 & -0.125 & 0.015 \\
\addlinespace
Worst worker quartile & -0.424 & -0.302 & -0.122 & 0.013 \\
Median workers & -0.257 & -0.172 & -0.085 & 0.014 \\
Best worker quartile & -0.137 & -0.086 & -0.051 & 0.029 \\
\midrule
\multicolumn{5}{l}{\textit{Panel C: Industry Quartiles, earnings effects}}\\
All worker types & -0.323 & -0.169 & -0.154 & 0.022 \\
\addlinespace
Worst worker quartile & -0.453 & -0.273 & -0.180 & 0.028 \\
Median workers & -0.282 & -0.184 & -0.098 & 0.019 \\
Best worker quartile & -0.126 & -0.088 & -0.037 & 0.036 \\
\addlinespace
\midrule
\multicolumn{5}{l}{\textbf{Mobility Estimates}}\\
\midrule
\multicolumn{5}{l}{\textit{Panel D: Location quartiles, effects on location mobility}}\\
All worker types & 0.018 & 0.007 & 0.011 & 0.002 \\
\addlinespace
Worst worker quartile & 0.012 & 0.011 & 0.001 & 0.002 \\
Median workers & 0.021 & 0.003 & 0.019 & 0.004 \\
Best worker quartile & 0.025 & 0.003 & 0.022 & 0.006 \\
\addlinespace
\midrule
\multicolumn{5}{l}{\textit{Panel E: Industry quartiles, effects on industry mobility}}\\
All worker types & 0.332 & 0.203 & 0.129 & 0.028 \\
\addlinespace
Worst worker quartile & 0.416 & 0.246 & 0.170 & 0.034 \\
Median workers & 0.313 & 0.218 & 0.095 & 0.028 \\
 Best worker quartile & 0.252 & 0.166 & 0.086 & 0.052 \\
\addlinespace
\bottomrule
  \multicolumn{5}{p{0.98\textwidth}}{\scriptsize \textit{Note}: Panels A--C display  displacement effects on earnings on year after displacements (calculated as displaced-control differences, "ATEs"). Panel A shows earnings estimates for the worst quartile of workers (Column 1) and the best quartile (Column 2), using the worker resiliency measure described above. Results for workers in all markets (interaction between industry and location), for workers in the worst (best) market quartiles, and for median markets (workers in the 3d or 4th market quintiles). Column 3--4 show the difference between Columns (1) and (2), and the standard error for the difference (clustered at the level of the pre-displacement establishment).
  Panel B shows earnings estimates for the best and the worst quartiles of locations using the partial-effects procedure described above. Panel C shows similar earnings measures when dividing by the best and worst industries. Both Panel B and C report estimates for all workers and when dividing workers by worker resiliency using the worker resiliency measures described above. Median workers are those in the 3d or 4th worker quintiles. Panel D shows displacement effects on location mobility (mobility defined as moving to another local labor market between one year before and three years after the displacement), and Panel E displacement effects on industry mobility (mobility defined as switching 1-digit industry between one year before and three years after the displacement). Panels D--E show results for the full sample of workers and when dividing the sample by worker resilience as in Panels B--C. Standard errors in Panel B and D clustered at the location level, and standard errors in Panel C and E clustered at the industry level. All estimates use the main data set. Rankings of workers and reshuffling done using the k-folds procedure.}
\end{tabularx}
\end{centering}
\end{table}


\subsection{Characteristics of Markets with Large Partial Effects}
Table \ref{Tab1_indreshuffle} describes the best and worst locations and industries (top and bottom quartiles of $\tau_l^L$ and $\tau_s^S$). Panel A shows location characteristics ($X^L$) for the \say{good} and \say{bad} locations as well as cross-quartile differences in $X^L$. Standard errors for differences are clustered at the location level. Locations with bottom-quartile partial effects ($\tau_l^L$) are in particular characterized by much lower population density. These locations also have high unemployment rates and a more concentrated industry structure dominated by declining industries and manufacturing jobs. 

Panel B shows similar results for industry characteristics. Industries with large predicted earnings losses for the average worker ($\tau_s^S$) are exclusively found in manufacturing, while industries with small predicted losses are in non-manufacturing sectors. Industries with very negative $\tau_s^S$-estimates also have higher wage premia, are less dynamic (lower churning and reallocation rates), and experience declining employment trends over both the short and the long run. All of these attributes are typical of manufacturing, but they are also related to effect heterogeneity if we only compare different manufacturing industries to each other, or if we only compare non-manufacturing industries to each other (see Table \ref{t:manuf}). 

\begin{table}[H]
\footnotesize
\caption{Characteristics for the worst and best manufacturing and non-manufacturing industries} \label{t:manuf}
\begin{centering}
\begin{tabularx}{\textwidth}{l zz zz zz}
\toprule
& \multicolumn{3}{c}{Manufacturing} & \multicolumn{3}{c}{Non-manufacturing} \\
\cline{2-4}  \cline{5-7} \\
&Worst Quartile&Best Quartile&Interquartile Difference &Worst Quartile&Best Quartile&Interquartile Difference\\
& (1) & (2) & (3) & (4) & (5) & (6)\\
\midrule
Employment trend & -0.284 & 0.062 & 0.346 & -0.048 & 0.348 & 0.395 \\
Employment cycle & -0.043 & -0.006 & 0.037 & -0.007 & 0.021 & 0.028 \\
Industry wage & 0.084 & 0.035 & -0.050 & 0.051 & -0.010 & -0.061 \\
Churning & 0.165 & 0.164 & -0.001 & 0.209 & 0.257 & 0.048 \\
Reallocation & 0.109 & 0.118 & 0.008 & 0.131 & 0.163 & 0.031 \\
Manufacturing & 1.000 & 1.000 & 0.000 & 0.000 & 0.000 & 0.000 \\
 Education, health, admin & 0.000 & 0.000 & 0.000 & 0.031 & 0.233 & 0.202 \\
\addlinespace
\bottomrule
\addlinespace
  \multicolumn{7}{p{0.98\textwidth}}{\scriptsize \textit{Note}: The table displays average industry characteristics for the best and the worst quartiles of manufacturing industries (Columns 1--3) and non-manufacturing industries (Columns 4--6). Both panels use the re-shuffling procedure described above. The industry characteristics are described in Section 3.2. Standard errors clustered at the industry level. All estimates use the main data set.}
\end{tabularx}
\end{centering}
\end{table}


Table \ref{t:worker_stats} describes the characteristics of the top and bottom partial worker-level quartiles ($\tau_i^I$) and compare these to differences across raw CATE quartiles.\footnote{In contrast to Figure \ref{fig5_interquartile}, we use raw non-standardized numbers in this table.} Overall, the results show relatively small differences between the partial and overall quartiles. This suggests that a very small share of the differences in treatment effects associated with worker characteristics arises because of correlations with market factors. The table reaffirms the strong relationships with age, schooling, and specific human capital. The most striking difference is related to gender. Females are highly over-represented in the lowest \textit{partial} quartile ($\tau_i^I$), despite not being in the lowest \textit{overall} CATE quartile. The (statistical) reason is that females are underrepresented in manufacturing where effects tend to be larger. There are also fewer immigrants in the best partial quartile than in the best overall CATE quartile, reflecting both an under-representation of immigrants in manufacturing and an over-representation of immigrants in metropolitan areas where effects tend to be more muted. Similarly, an over-representation of STEM graduates in the most resilient partial quartile reflects that many engineers work in manufacturing, but still suffer from relatively modest effects.


\begin{table}[H]
\caption{Worker characteristics for resilient and non-resilient workers}\label{t:worker_stats}
\footnotesize
\begin{centering}
\begin{tabularx}{\textwidth}{l zz zz zz}
\toprule
& \multicolumn{3}{c}{Workers ranked by} & \multicolumn{3}{c}{Workers ranked by } \\
& \multicolumn{3}{c}{\emph{partial} CATE} & \multicolumn{3}{c}{CATE} \\
\cline{2-4} \cline{5-7}
& (1) & (2) & (3) & (4) & (5) & (6) \\
&Worst Worker Quartile & Best Worker Quartile & Difference [1]-[2]  &Worst Worker Quartile & Best Worker Quartile  & Difference [1]-[2] \\ 
\midrule
 \multicolumn{5}{l}{\emph{Demographics}} \\
Age & 52.331 & 35.626 & 16.705                                               & 50.736 & 36.136 & 14.600 \\
Female & 0.409 & 0.308 & 0.101                                               &0.377 & 0.363 & 0.015  \\             
Born abroad & 0.167 & 0.046 & 0.121                                          &0.164 & 0.079 & 0.085 \\              
Two foreign born parents & 0.025 & 0.033 & -0.008                           &0.025 & 0.038 & -0.013   \\           
Native, not in birth country & 0.216 & 0.308 & -0.093                        &0.202 & 0.299 & -0.097   \\           
& \\
\multicolumn{5}{l}{\emph{Family}} \\
Married & 0.641 & 0.542 & 0.099                                              &0.634 & 0.527 & 0.107 \\              
Divorced & 0.159 & 0.048 & 0.112                                             &0.146 & 0.059 & 0.086 \\              
Children & 0.330 & 0.918 & -0.588                                            &0.405 & 0.868 & -0.463   \\           
School-aged children & 0.240 & 0.506 & -0.266                                &0.293 & 0.486 & -0.194   \\           
Share household inc. & 0.710 & 0.797 & -0.087                                &0.718 & 0.788 & -0.070   \\           
Past mobility & 0.093 & 0.345 & -0.252                                       &0.104 & 0.330 & -0.226   \\           
& \\
\multicolumn{5}{l}{\emph{Human capital}} \\
Schooling (years of education) & 10.132 & 12.907 & -2.776                    & 10.145 & 12.648 & -2.503 \\          
Experience & 9.360 & 8.649 & 0.710                                           &9.310 & 8.557 & 0.753 \\              
Tenure & 6.702 & 5.278 & 1.423                                               &6.771 & 5.189 & 1.583 \\              
Industry tenure & 7.934 & 6.533 & 1.401                                      &7.938 & 6.484 & 1.454 \\              
Earnings (rank) & 0.364 & 0.654 & -0.290                                     &0.373 & 0.595 & -0.222   \\           
STEM education & 0.034 & 0.247 & -0.214                                      &0.035 & 0.202 & -0.167   \\           
Education in licensed field & 0.059 & 0.085 & -0.026                         &0.049 & 0.102 & -0.053   \\           
Education specificity & 0.431 & 0.522 & -0.092                               &0.423 & 0.523 & -0.100   \\      
\bottomrule
  \multicolumn{7}{p{0.98\textwidth}}{\scriptsize \textit{Note}: The table shows average worker characteristics differ for the best and worst quartiles of workers. Columns 1--3 rank workers by CATE and Columns 4--6 by the partial worker resiliency measure described above. Columns 2--3 and 4--5 display averages within the respective quartiles. Column (3) shows differences between Columns (1) and (2). The worker characteristics are described in Section 3.2. All sample statistics are for the main data set. Rankings of workers and reshuffling done using the k-folds procedure.}
\end{tabularx}
\end{centering}
\end{table}

\clearpage


\end{document}